\documentclass{article}
\usepackage{tabularx}

\usepackage{listings}
\usepackage{microtype}
\usepackage{graphicx}
\usepackage{subcaption}
\usepackage{enumitem}
\usepackage{booktabs} 

\usepackage{hyperref}

\usepackage[preprint]{icml2026}

\usepackage{amsmath}
\usepackage{amssymb}
\usepackage{mathtools}
\usepackage{amsthm}

\usepackage[capitalize,noabbrev]{cleveref}

\theoremstyle{plain}

\theoremstyle{definition}

\theoremstyle{remark}

\usepackage[textsize=tiny]{todonotes}

\icmltitlerunning{IndustryCode: A Benchmark for Industry Code Generation}
\begin{document}

\twocolumn[
  \icmltitle{IndustryCode: A Benchmark for Industry Code Generation}



  \icmlsetsymbol{equal}{*}

  \begin{icmlauthorlist}
    \icmlauthor{Puyu Zeng}{sjtu}
    \icmlauthor{Zhaoxi Wang}{sjtu}
    \icmlauthor{Zhixu Duan}{sjtu}
    \icmlauthor{Liang Feng}{sjtu}
    \icmlauthor{Shaobo Wang}{alibaba,sjtu}
    \icmlauthor{Cunxiang Wang}{tsinghua}
    \icmlauthor{Jinghang Wang}{alibaba}
    \icmlauthor{Bing Zhao}{alibaba}
    \icmlauthor{HU WEI}{alibaba}
    \icmlauthor{Linfeng Zhang}{sjtu}
  \end{icmlauthorlist}

  \icmlaffiliation{sjtu}{Shanghai Jiao Tong University, Shanghai, China}
  \icmlaffiliation{alibaba}{Alibaba Group, Hangzhou, China}
  \icmlaffiliation{tsinghua}{Tsinghua University, Beijing, China}

  \icmlcorrespondingauthor{Linfeng Zhang}{zhanglinfeng@sjtu.edu.cn}

  \icmlkeywords{Machine Learning, Code Generation, Benchmark}

  \vskip 0.3in
]



\printAffiliationsAndNotice{}  

\begin{abstract}
  Code generation and comprehension by Large Language Models (LLMs) have emerged as core drivers of industrial intelligence and decision optimization, finding widespread application in fields such as finance, automation, and aerospace.
  Although recent advancements have demonstrated the remarkable potential of LLMs in general code generation, existing benchmarks are mainly confined to single domains and languages.
  Consequently, they fail to effectively evaluate the generalization capabilities required for real-world industrial applications or to reflect the coding proficiency demanded by complex industrial scenarios.
  To bridge this gap, we introduce IndustryCode, the first comprehensive benchmark designed to span multiple industrial domains and programming languages.
  IndustryCode comprises 579 sub-problems derived from 125 primary industrial challenges, accompanied by rigorous problem descriptions and test cases. 
  It covers a wide range of fields, including finance, automation, aerospace, and remote sensing—and incorporates diverse programming languages such as MATLAB, Python, C++, and Stata. 
  In our evaluation, the top-performing model, Claude 4.5 Opus, achieved an overall accuracy of 68.1\% on sub-problems and 42.5\% main problems. The benchmark dataset and automated evaluation code will be made publicly available upon acceptance.
\end{abstract}
\begin{figure*}[t] 
  \begin{center}
    \centerline{\includegraphics[width=0.85\textwidth]{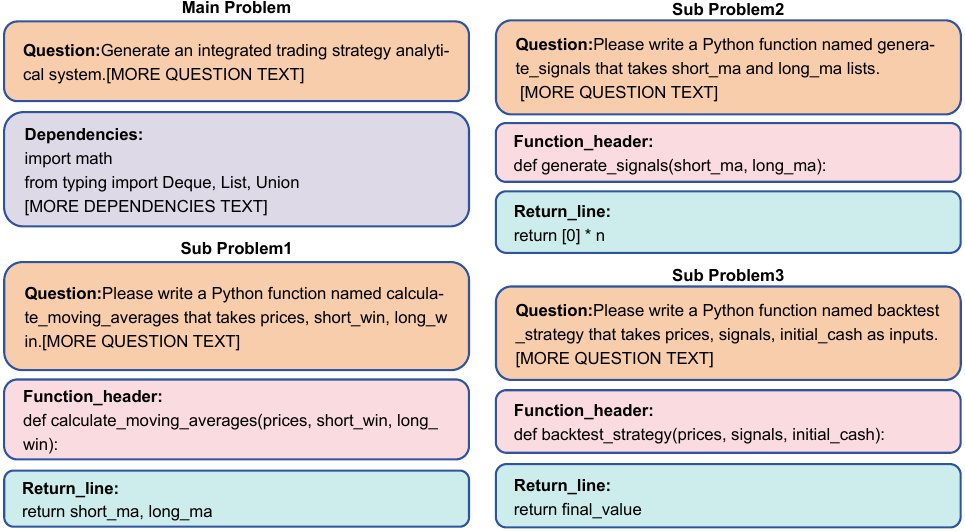}}
    \vspace{-0.1cm}
    \caption{
      \textbf{Hierarchical decomposition of an IndustryCode task.} 
      A complex Main Problem is factorized into multiple modular Sub-problems 
      to simulate real-world development workflows. Each component includes detailed 
      functional requirements, necessary library dependencies, and precise function signatures.
    }
    \label{icml-historical}
  \end{center}
  \vskip -0.3in
\end{figure*}
\section{Introduction}
The breakthrough advancements of Large Language Models (LLMs) in code generation and program comprehension have triggered a profound paradigm shift in programming~\cite{peng2023impact, openai2023gpt4, feng2020codebert, wang2021codet5, li2022competition}. Modern Code LLMs not only demonstrate exceptional proficiency in syntactic correctness and logical coherence but have also emerged as pivotal engines driving productivity surges across diverse fields. In the field of software engineering, next-generation AI-native development environments, represented by Claude Code, Cursor, and Trae, are significantly shortening development lifecycle by automating complex tasks ranging from code completion to test-driven development (TDD) and system refactoring. Concurrently, in the natural sciences, LLMs are democratizing access to scientific computing~\cite{boiko2023autonomous,zheng2025large}, enabling researchers in fields such as physics and biology to construct experimental models via natural language interaction. These tools effectively lower the technical barriers and accelerate the pace of scientific discovery.

Existing mainstream code evaluation benchmarks mainly focus on the domain of General Software Development~\cite{chen2021evaluating,austin2021program,jain2024livecodebench,yu2024humaneval}. For instance, SWE-bench~\cite{jimenez2024swebench} assesses capabilities in software maintenance by simulating real-world GitHub issue resolution, while V-GameGym~\cite{zhang2025vgamegym} focuses on logic generation and interaction in game environments. However, these benchmarks largely reflect the typical paradigms of internet or consumer-level software and are often confined to a single domain or programming language. Consequently, they fail to adequately address the rigorous demands for numerical precision, linguistic diversity, and domain-specific knowledge inherent to vertical sectors such as industrial automation, aerospace, and quantitative finance~\cite{zhang2025xfinbench}. The complexity of these industrial scenarios brings a series of unique challenges:

\begin{itemize}
  \vspace{-0.1cm}
  \item \textbf{Domain Specificity:} Industrial coding involves strict physical constraints and specialized business logic (e.g., orbital calculations in aerospace or econometric assumptions in finance), which are often absent in generic benchmarks.~\cite{ishida2024langprop, rink2018cfdlang, pandey2025openfoamgpt, haag2025training,song2025code,he2025using}
  \vspace{-0.1cm}
  \item \textbf{Linguistic Diversity:} The industry heavily relies on specialized languages such as MATLAB for scientific computing or Stata for statistical analysis. Existing benchmarks lack the capacity to evaluate cross-lingual generalization.~\cite{articlematlab,o2025industry}
  \vspace{-0.1cm}
  \item \textbf{Task Complexity:} Real-world industrial problems require decomposing high-level requirements into interconnected sub-tasks, a capability not adequately tested by simple code snippets. Thus, current benchmarks fall short of effectively measuring the generalization ability of LLMs in genuine industrial contexts, creating a gap between benchmark scores and practical utility.
\end{itemize}

To bridge this gap and establish a gold standard for industrial-grade code evaluation, we introduce IndustryCode, the first comprehensive benchmark encompassing diverse industrial domains and programming languages. The dataset comprises 125 Main Problems, each hierarchically decomposed into constituent Sub-problems, aggregating to a total of 579 tasks. Each instance is paired with a specific problem\_description and test\_cases. To resolve a Main Problem, Large Language Models (LLMs) are required to implement modular components for the sub-problems and subsequently synthesize them into a unified solution. An illustrative example is presented in Figure 1. Designed to rigorously mirror authentic production environments, IndustryCode is characterized by the following key attributes:

\begin{table*}[t] 
  \caption{\textbf{Overview of IndustryCode fields and subfields.} The table lists the industrial domains covered by each programming language, with the number of problems in parentheses.}
  \label{fields-table}
  \begin{center}
    \begin{small}
      \begin{tabular}{lp{13cm}}
        \toprule
        \textbf{Programming Language} & \textbf{Subfields} \\
        \midrule
        Python & Chemical Engineering (14), Advanced Manufacturing (12), Finance and Business (11), Automation and Control Engineering (9), Artificial Intelligence (9),  Electronics and Communication (5), Biomedical Engineering (4), Information Technology (4), Transportation and Logistics (3)\\
        \midrule
        C++ & Construction (10), Information Technology (7), Mining (6), Machinery Manufacturing (5), Semiconductors and Microelectronics (4), Transportation and Logistics (4) \\
        \midrule
        Matlab & Operations Research (7), Genetic Computation (7), Aerospace Engineering (2), Naval Architecture and Ocean Engineering (1) \\
        \midrule
        Stata & Finance and Business (3) \\
        \bottomrule
      \end{tabular}
    \end{small}
  \end{center}

  \vskip -0.1in
  
\end{table*}

\begin{itemize}
  \vspace{-0.2cm}
  \item \textbf{Cross-Domain Coverage:} Encompassing high-barrier sectors, including Finance, Automation, Aerospace, and Remote Sensing, and so on.
  \vspace{-0.2cm}
  \item \textbf{High-Quality Data Construction:} Leveraging a human-AI collaborative pipeline involving GPT-5, we rigorously corrected and annotated the raw data, resulting in a curated set of 579 high-quality problems.
  
  \vspace{-0.15cm}
  \item \textbf{Authentic Industrial Source:} The dataset is derived exclusively from the actual production code of industry practitioners, ensuring that IndustryCode maintains high fidelity to real-world engineering practices.
  \vspace{-0.2cm}
  \item \textbf{ Decontamination:} To mitigate the risk of data contamination, all problems underwent meticulous filtration and adversarial manual revision, differentiating them from public corpora used in LLM pre-training.
  \vspace{-0.2cm}
  \item \textbf{Multi-Lingual Support:} Incorporating diverse languages such as MATLAB, Python, and Stata, reflecting the varied technology stacks used in the industry.
  \vspace{-0.2cm}
  \item \textbf{Hierarchical Problem Structure:} Comprising 579 sub-problems decomposed from 125 primary industrial challenges, equipped with detailed descriptions and rigorous test cases to evaluate reasoning and implementation capabilities.

\end{itemize}

We conducted extensive evaluations of state-of-the-art LLMs using IndustryCode.
Empirical results demonstrate that while models show promise, challenges persist in achieving industrial-grade reliability.
The top-performing model, Claude Opus 4.5, achieved an accuracy of 68.1\%, indicating substantial progress in understanding complex industrial logic while highlighting areas for future improvement.
We believe IndustryCode not only demonstrates the capabilities of LLMs as useful industrial assistants but also provides critical insights and a foundational evaluation framework for the future development of Industrial AI.

\section{IndustryCode}

This section elaborates on the design principles and annotation pipeline 
of IndustryCode. Specifically, it covers: 
the collection of research-grade coding problems from diverse industrial domains, 
the strategy for decomposing complex industrial challenges into executable sub-tasks, the annotation process tailored to the specific characteristics of industrial code, and
the evaluation framework.

\subsection{Diverse and Realistic Industrial Coding Problems}

IndustryCode gathers highly realistic and generalizable industrial coding challenges,
encompassing key areas such as industrial automation, aerospace, finance,
and artificial intelligence. Spanning twenty sub-domains and supporting four programming languages,
this diversity ensures a comprehensive representation of scientific and engineering disciplines 
where extensive code development is indispensable.

The core data of IndustryCode is derived from authentic codebases used by industry professionals 
in real-world production environments. Although the theoretical correctness of the underlying algorithms has been validated in numerous academic publications, the source code mainly consists of proprietary, closed-source implementations, retaining the defects and noise typical of actual development. To mitigate potential data contamination and memory risks in LLMs, we manually reconstructed the code and injected additional complexity, ensuring that all modifications passed strict functional testing. This provides a reliable foundation for precisely evaluating model generalization boundaries and coding capabilities within specific industrial contexts. Ultimately, IndustryCode contains 125 core industrial problems, hierarchically decomposed into 579 sub-problems.

Table 1 presents the sub-domains covered by IndustryCode and the distribution of main problems across each domain. We reserved 19 main problems (80 sub-problems) as the development split and used the remaining 106 main problems (499 sub-problems) as the test data.

\begin{figure}[ht]
    \vskip 0.2in
    \begin{center}
        \begin{subfigure}[b]{0.48\columnwidth}
            \centering
            \includegraphics[width=\linewidth]{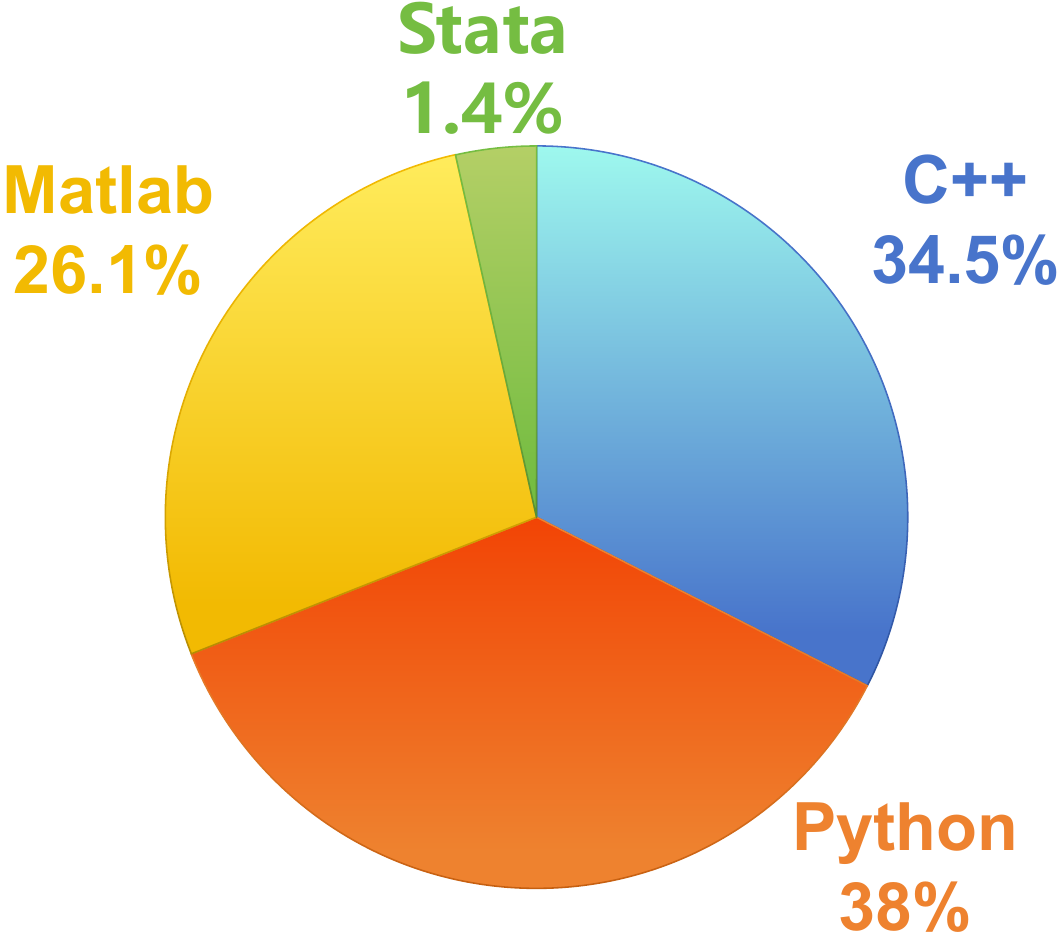}
            \caption{} 
            \label{fig:sub-a}
        \end{subfigure}
        \hfill
        \begin{subfigure}[b]{0.48\columnwidth}
            \centering
            \includegraphics[width=\linewidth]{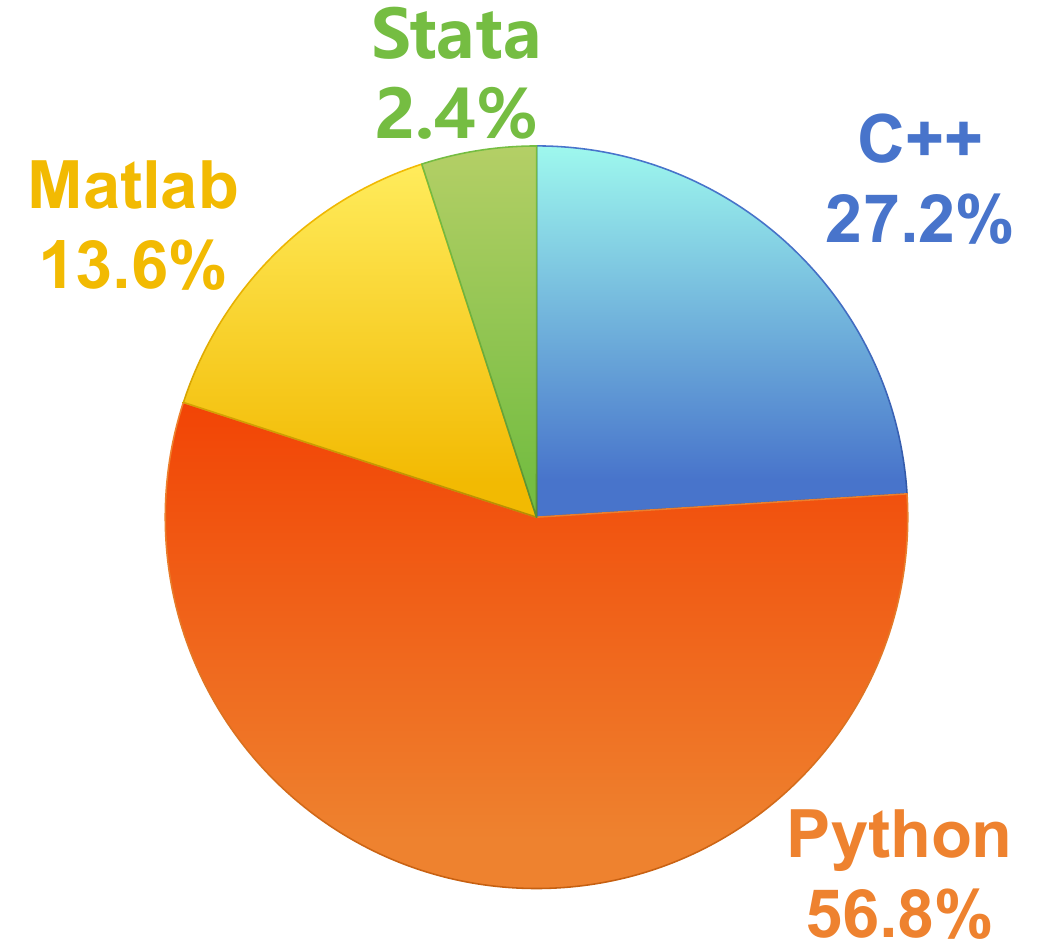}
            \caption{} 
            \label{fig:sub-b}
        \end{subfigure}
        
        \caption{\textbf{Task distribution across programming languages in IndustryCode.} The pie charts illustrate the proportional composition of the dataset. (a) Breakdown of sub-problems by languages. (b) Breakdown of main problems by languages.}
        \label{icml-historical}
    \end{center}
\end{figure}

\subsection{A Main Problem with Multiple Sub-problems}

In real-world industrial scenarios, practitioners typically decompose complex engineering projects 
into smaller, manageable sub-modules. They implement functionally independent functions for each module 
and integrate them to construct the final system. In constructing this dataset, we adopted this 
structured approach to ensure alignment with natural engineering practices.

\textbf{Main Problem:}
Each main problem corresponds to a complete industrial-level engineering project. 
It delineates the overall objectives, global inputs, and expected outputs, providing 
high-level guidance for subsequent engineering decomposition and implementation. 
These main problems can be further broken down into a series of logically coherent and functionally focused sub-problems. 
This hierarchical structure reflects the whole-to-part system development paradigm found in real-world industrial scenarios, serving as the foundational unit for evaluating LLMs' capabilities in task understanding and code organization within a comprehensive engineering workflow.
\begin{figure}[ht]
  \vskip 0.2in
  \begin{center}
    \centerline{\includegraphics[width=\columnwidth]{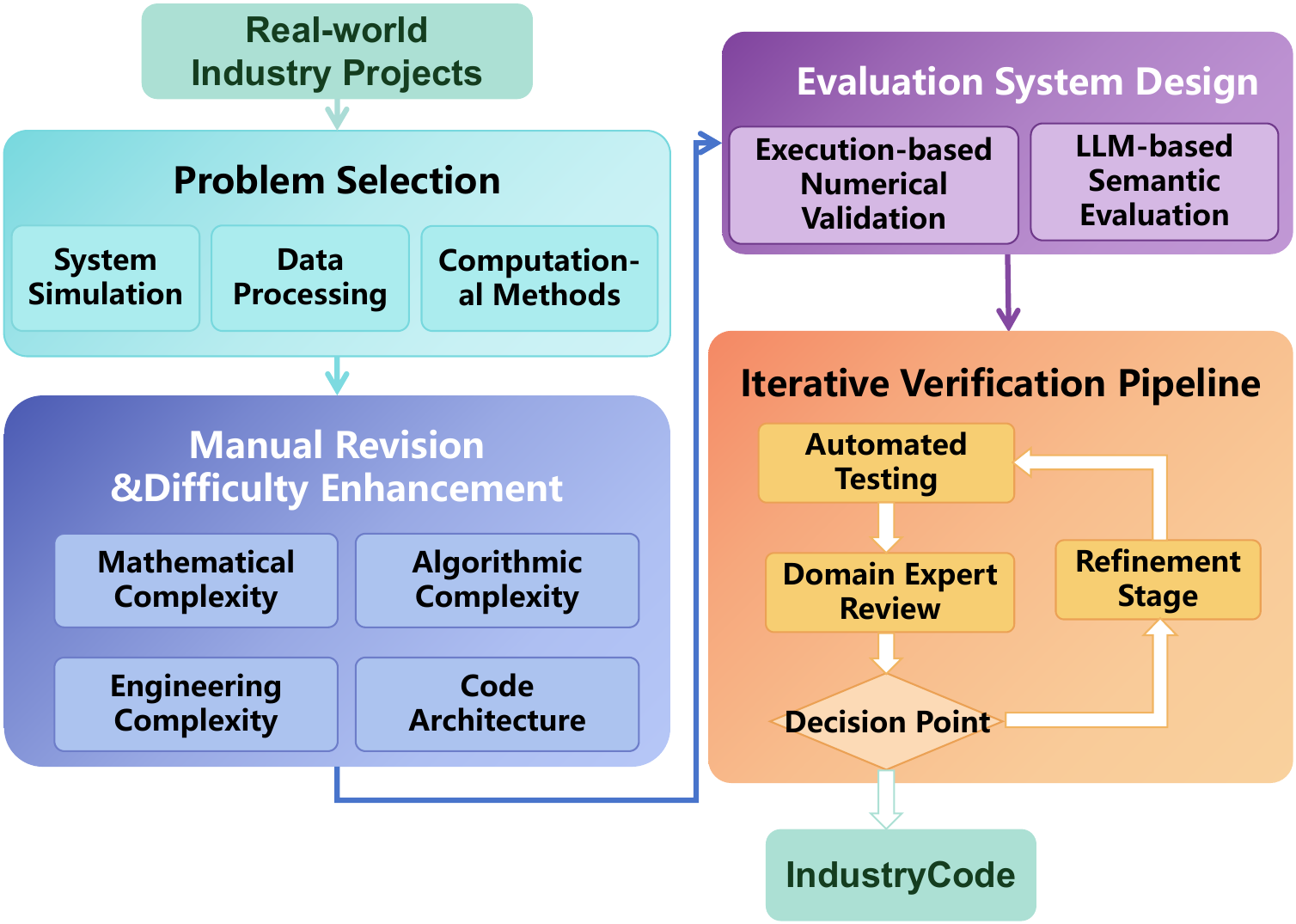}}
    \caption{
      \textbf{Data Annotation flowchart}
    }
    \vspace{-0.4cm}
    \label{icml-historical}
  \end{center}
\end{figure}

\textbf{Sub-problem:}
Sub-problems are derived from the main problem through hierarchical decomposition, 
aiming to break down complex engineering projects into sub-modules that are functionally cohesive, clearly delimited, and easy to implement. 
Each sub-problem includes detailed input/output specifications and functional descriptions to facilitate accurate code generation. 
This layered structure not only significantly reduces the complexity of industrial-level development, but also provides a granular basis 
for evaluating LLMs' capabilities in fine-grained task understanding, logical organization, and coding accuracy.


\subsection{Data Annotation}

The construction of IndustryCode follows a systematic engineering workflow, including the following four phases:
\begin{itemize}

\item \textbf{Problem Selection:} Guided by real-world industrial demands and research objectives, we selecte representative coding task topics across multiple domains.
\vspace{-0.1cm}
\item \textbf{Manual Revision and Difficulty Enhancement:} To mitigate evaluation bias caused by data contamination, we manually reformulate the original problems. Furthermore, we enhance the difficulty by incorporating rigorous mathematical logic, complex engineering implementations, and specific design patterns.
\vspace{-0.1cm}
\item \textbf{Evaluation System Design:} We develop comprehensive numerical test cases and functional verification samples for each problem, ensuring that the correctness and completeness of the generated code are assessed in an objective and reproducible manner.
\vspace{-0.1cm}
\item \textbf{Iterative Verification:} We employ a hybrid iterative mechanism combining Human Review and LLM-assisted Analysis to further refine and optimize the quality of problem design.

\end{itemize}

\begin{figure}[ht]
  \vskip 0.2in
  \begin{center}
    \centerline{\includegraphics[width=\columnwidth]{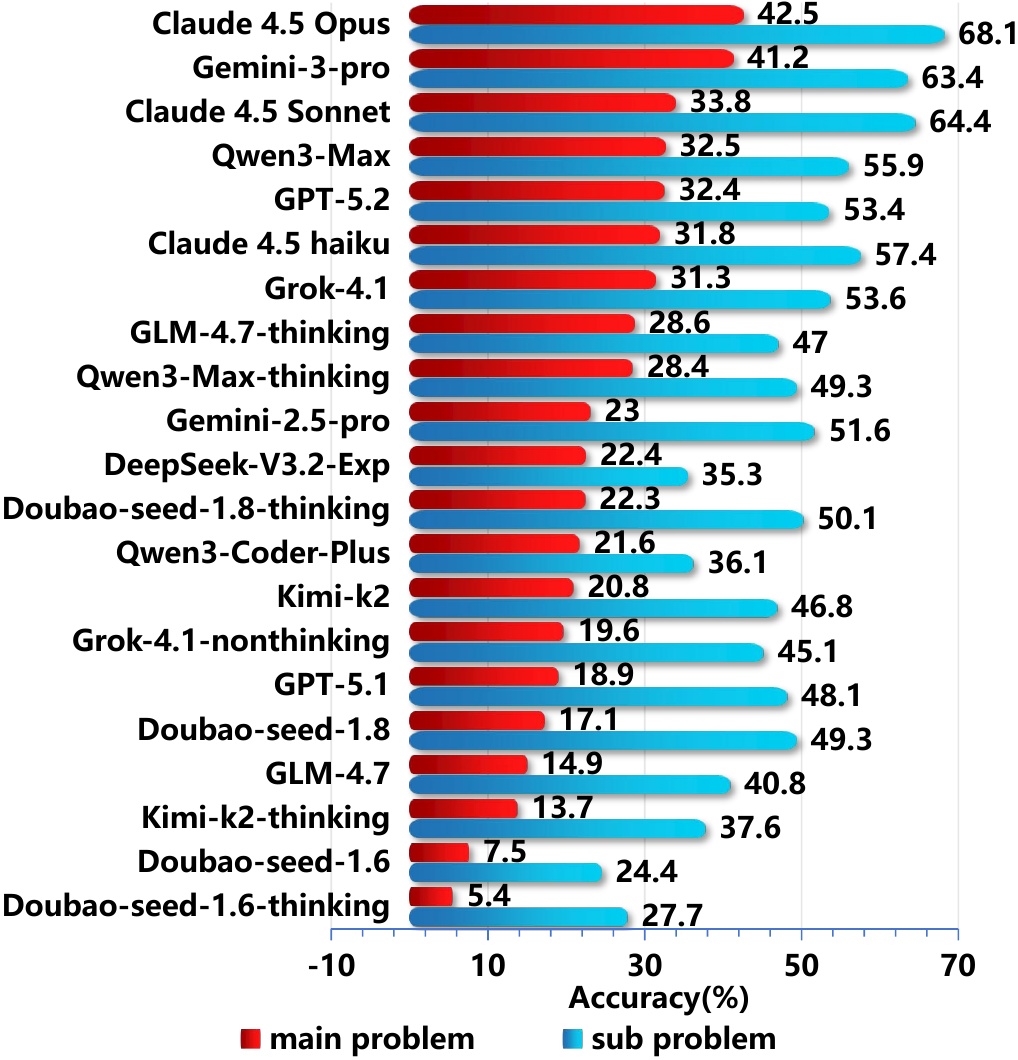}}
    \caption{
      \textbf{Performance comparison on main problems and sub-problems.}The observed trends indicate that strong foundational capabilities in sub-problems generally translate to better performance in  main problems.
    }
    \vspace{-0.4cm}
    \label{icml-historical}
  \end{center}
\end{figure}

\subsubsection{Problem Selection}
To capture the diverse coding demands of the engineering lifecycle ranging from simulation to numerical calculation, we construct IndustryCode with problems requiring deep domain expertise and reasoning capabilities. This design enables a rigorous assessment of model proficiency in authentic industrial contexts by encompassing the breadth and depth of critical applications. The benchmark specifically targets system simulation for modeling complex behaviors, data processing essential for project execution, and computational methods for obtaining key values through algorithms rather than raw data observation.

\begin{figure}[t]
  \vskip 0.2in
  \begin{center}
    \centerline{\includegraphics[width=0.7\columnwidth]{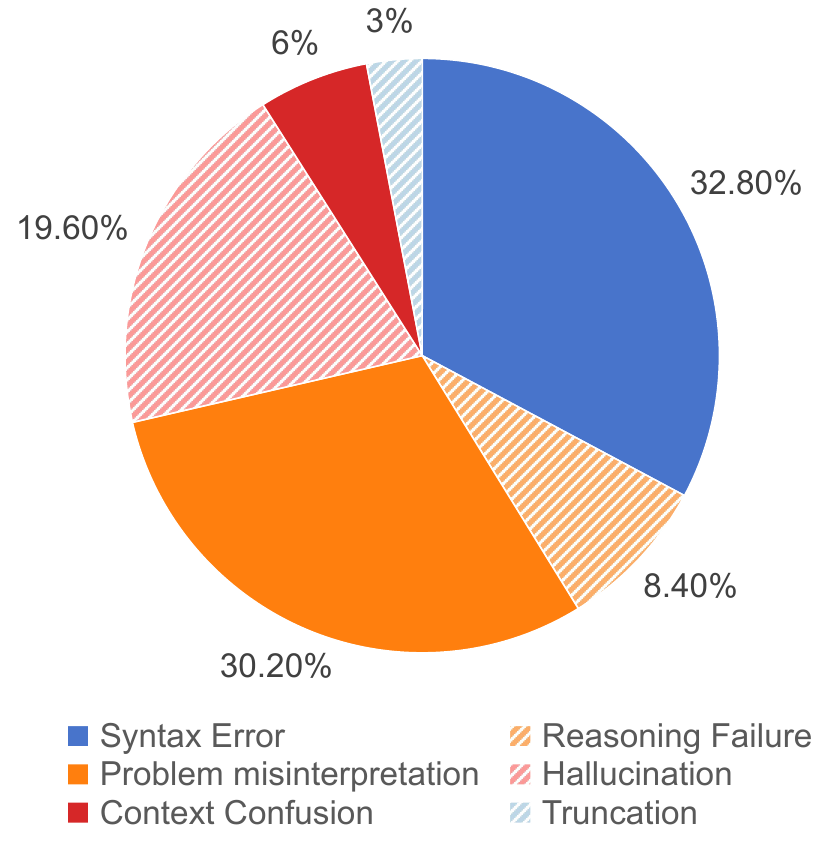}}
    \caption{
      \textbf{Distribution of failure cases in IndustryCode.} 
    }
    \vspace{-0.4cm}
    \label{icml-historical}
  \end{center}
\end{figure}

\subsubsection{Manual Revision and Difficulty Enhancement}
In constructing IndustryCode, our primary objective is to distinguish true engineering reasoning from rote memorization of training corpora. To achieve this, we conduct rigorous manual revisions and adversarial hardening on the raw problems extracted from actual industrial projects, mainly focusing on four strategic dimensions:

\begin{itemize}

\item \textbf{Mathematical Complexity:} We introduce complex constraints to significantly raise the standard of mathematical rigor. These constraints require high-order methods in numerical linear algebra and non-linear optimization to ensure excellent numerical stability.
\vspace{-0.1cm}
\item \textbf{Algorithmic Complexity:} We replace standard implementations with complex algorithm logic to test the models' ability to handle higher computational complexity and logical density.
\vspace{-0.1cm}
\item \textbf{Engineering Complexity:} To replicate the complexity of real-world engineering, we increase the degree of inter-module coupling and embedded comprehensive exception handling mechanisms for edge cases.
\vspace{-0.1cm}
\item \textbf{Code Architecture:} In addition to functional correctness, we also require the adoption of specific design patterns to strictly adhere to industrial standards for maintainability and modularity.

\end{itemize}

\subsubsection{Evaluation System Design:}

To ensure a comprehensive evaluation of code generation quality, our evaluation suite includes two key components:

\begin{enumerate}[label=(\arabic*)]
\vspace{-0.2cm}
\item \textbf{Execution-based Numerical Validation:} To meet the strict precision requirements of industrial code, we designed a set of input-output (I/O) pairs for dynamic testing. This component executes the generated code to verify whether the outputs align precisely with the true values, thereby ensuring correctness at the numerical computation level.
\vspace{-0.2cm}
\item \textbf{LLM-based Semantic Evaluation (LLM-Judge):} Complementing numerical metrics, we introduce an LLM-Judge framework equipped with rigorously designed prompts. This framework conducts a comparative analysis between the generated and reference code. It not only evaluates functional equivalence but also identifies syntactic and structural discrepancies, ensuring the code adheres to industrial standards.
\end{enumerate}

\begin{figure}[t]
  \begin{center}
    \centerline{\includegraphics[width=\columnwidth]{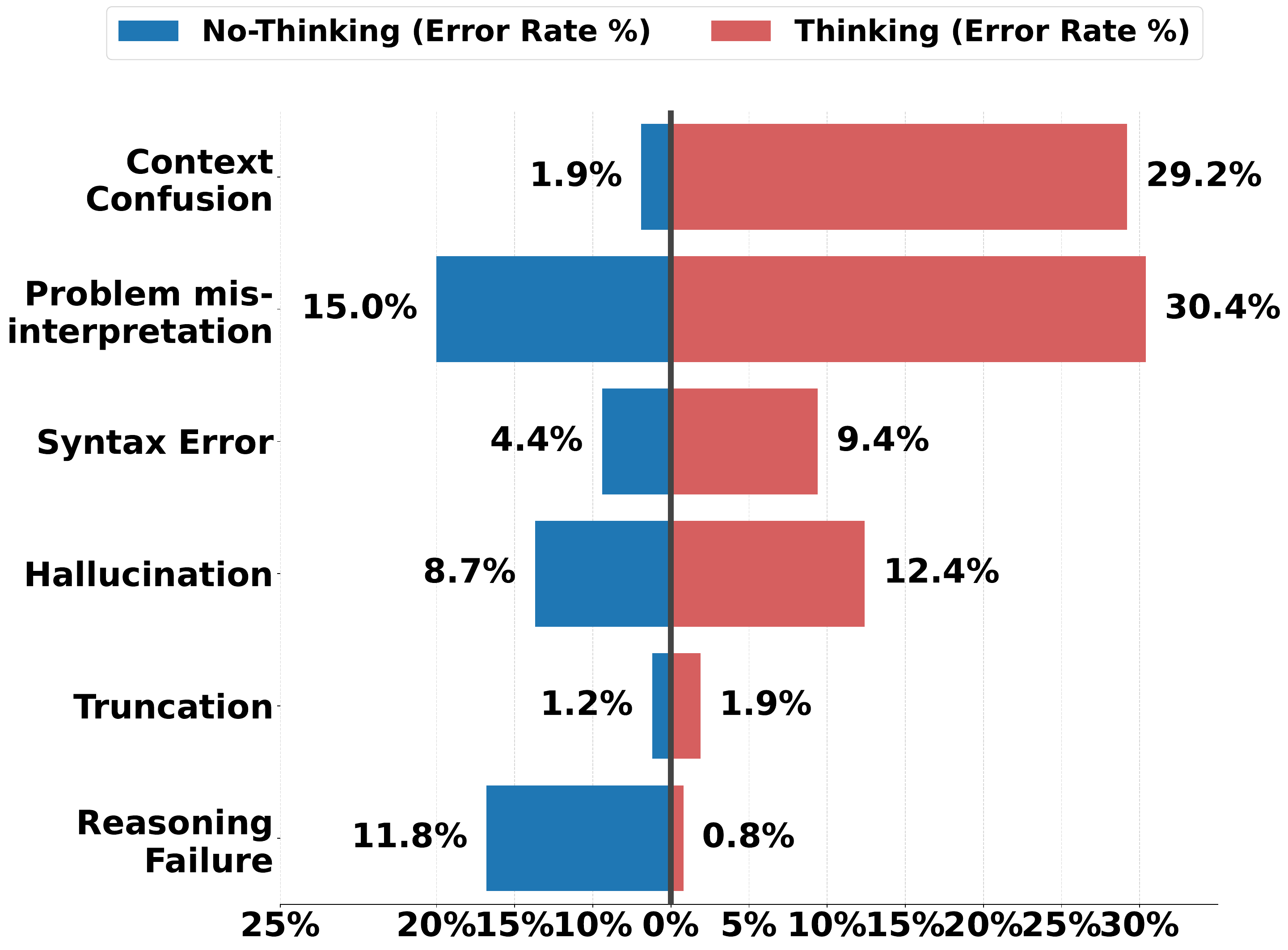}}
    \caption{
      \textbf{Impact of Thinking Mode on the distribution of failure modes.}The results show a marked reduction in reasoning errors, counterbalanced by a significant surge in context confusion and misunderstandings.
    }
    \vspace{-0.5cm}
    \label{icml-historical}
  \end{center}
\end{figure}

\subsubsection{Iterative Verification:}
To ensure that problems remains mathematical solvable and industrially consistent after manual difficulty enhancement, we established a three-stage iterative validation pipeline. We first conduct automated tests on all the code samples within pre-configured industrial-grade runtime environments such as MATLAB and Stata to filter out syntax errors or missing dependencies that might be introduced during refactoring. After this stage, the samples undergo a second round of review by domain experts to verify whether the adversarial hardening has affected adherence to mathematical laws or industry standards. We then flagged and rejected any cases that violated industrial norms.  Finally,we conducted a closed-loop refinement process, iteratively fine-tuning defective problems until all 579 sub-problems consistently produced results within the engineering tolerance of the ground truth values.






\newcolumntype{Y}{>{\centering\arraybackslash}X}

\begin{table*}[t]
  \caption{\textbf{Comparison of Proprietary and Open-Weights LLMs on IndustryCode.} The table reports Pass@1 accuracy (\%) across four industrial programming domains plus the overall performance.The Overall column represents the average score of the four languages.}
  \label{model-comparison-table}
  \begin{center}
    \begin{small}
      \begin{tabularx}{\textwidth}{lYYYYYYYYYY}
        \toprule
        & \multicolumn{2}{c}{\textbf{Python}} & \multicolumn{2}{c}{\textbf{C++}} & \multicolumn{2}{c}{\textbf{Matlab}} & \multicolumn{2}{c}{\textbf{Stata}} & \multicolumn{2}{c}{\textbf{Overall}} \\
        \cmidrule(lr){2-3} \cmidrule(lr){4-5} \cmidrule(lr){6-7} \cmidrule(lr){8-9} \cmidrule(lr){10-11}
        \textbf{Models} & \textbf{Sub} & \textbf{Main} & \textbf{Sub} & \textbf{Main} & \textbf{Sub} & \textbf{Main} & \textbf{Sub} & \textbf{Main} & \textbf{Sub} & \textbf{Main} \\
        \midrule
        
        \textit{Proprietary Models} & & & & & & & & & & \\
        \hspace{1em}Claude 4.5 Opus      & \textbf{61.8} & \textbf{36.5} & \textbf{70.9} & \textbf{52.0} & 72.8 & 31.3 & \textbf{66.7} & 50.0 & \textbf{68.1} & \textbf{42.5} \\
        \hspace{1em}Claude 4.5 Sonnet    & 58.6 & \textbf{36.5} & 58.0 & 36.0 & 74.3 & \textbf{37.5} & 66.7 & 50.0 & 64.4 & 33.8 \\
        \hspace{1em}Claude 4.5 haiku     & 53.4 & 28.6 & 61.7 & 36.0 & 64.3 & 12.5 & 50.0 & 50.0 & 57.4 & 31.8 \\
        \hspace{1em}GPT-5.2              & 48.5 & 27.0 & 67.9 & 40.0 & 47.1 & 12.5 & 50.0 & 50.0    & 53.4 & 32.4 \\
        \hspace{1em}GPT-5.1              & 48.5 & 27.0 & 63.5 & 36.0 & 47.1 & 12.5 & 33.3 & 0.0    & 48.1 & 18.9 \\
        \hspace{1em}Gemini-3-pro         & 59.7 & 41.3 & 66.0 & 36.0 & \textbf{77.9} & 37.5 & 50.0 & 50.0 & 63.4 &  41.2 \\
        \hspace{1em}Gemini-2.5-pro       & 54.5 & 34.9 & 58.6 & 32.0 & 60.0 & 25.0 & 33.3    & 0.0   & 51.6 & 23.0 \\           
        \hspace{1em}Doubao-seed-1.6      & 38.4 & 17.5 & 12.9 & 0.0    & 46.4 & 12.5 & 0.0 & 0.0    & 24.4 & 7.5 \\
        \hspace{1em}Doubao-seed-1.6-thinking& 44.5 & 17.4 & 17.3 & 4.0  & 15.7 & 0.0    & 33.3 & 0.0    & 27.7 & 5.4 \\
        \hspace{1em}Doubao-seed-1.8      & 45.5 & 25.4 & 51.2 & 24.0 & 67.1 & 18.8 & 33.3 & 0.0    & 49.3 & 17.1 \\
        \hspace{1em}Doubao-seed-1.8-thinking& 53.4 & 30.2 & 54.3 & 40.0 & 59.3 & 18.8 & 33.3 & 0.0    & 50.1 & 22.3 \\    
        \hspace{1em}Grok-4.1-nonthinking & 47.6 & 25.4 & 48.1 & 28.0 & 51.4 & 25.0 & 33.3    & 0.0   & 45.1 & 19.6 \\ 
        \hspace{1em}Grok-4.1             & 42.4 & 22.2 & 51.2 & 28.0 & 54.2 & 25.0 & 66.7    & 50.0   & 53.6 & 31.3 \\ 
        
        \midrule
        \textit{Open Models} & & & & & & & & & & \\
        \hspace{1em}DeepSeek-V3.2-Exp    & 49.5 & 25.4 & 13.6 & 8.0  & 27.9 & 6.3    & 50.0 & 50.0 & 35.3 & 22.4 \\
        \hspace{1em}Qwen3-Coder-Plus     & 50.0 & 31.7 & 54.3 & 36.0 & 40.0 & 18.8 & 0.0    & 0.0    & 36.1 & 21.6 \\
        \hspace{1em}Qwen3-Max            & 56.6 & 31.7 & 70.4 & 36.0 & 46.4 & 12.5 & 50.0 & 50.0    & 55.9 & 32.5 \\
        \hspace{1em}Qwen3-Max-thinking   & 38.2 & 20.6 & 59.9 & 24.0 & 49.3 & 18.8 & 50.0 & 50.0    & 49.3 & 28.4 \\
        
        \hspace{1em}GLM-4.7              & 40.3 & 28.6 & 44.4 & 12.0 & 45.0 & 18.8 & 33.3 & 0.0    & 40.8 & 14.9 \\
        \hspace{1em}GLM-4.7-thinking     & 50.3 & 30.2 & 46.3 & 28.0 & 41.4 & 6.3  & 50.0 & 50.0 & 47 & 28.6 \\
        \hspace{1em}Kimi-k2              & 48.7 & 22.2 & 54.3 & 36.0 & 50.7 & 25.0 & 33.3 & 0.0    & 46.8 & 20.8 \\
        \hspace{1em}Kimi-k2-thinking     & 51.8 & 23.8 & 25.3 & 12.0 & 40.0 & 18.8 & 33.3 & 0.0    & 37.6 & 13.7 \\

        \bottomrule
      \end{tabularx}
    \end{small}
  \end{center}
  \vskip -0.1in
\end{table*}

\section{Experiments}
\textbf{Prompts:} To rigorously evaluate the intrinsic coding capabilities of Large Language Models in IndustryCode, we adopt a unified Zero-shot Prompting strategy. Although prompt templates are tailored to the syntactic characteristics of specific programming languages, we maintain strict structural consistency across all evaluated models, languages, and domains. These prompts are only used to describe task specifications, without any few-shot demonstrations.

It is crucial that we implement a cumulative context mechanism to simulate the real industrial continuous development workflow. The prompts provided to the model comprise three core components: (1) a global description of the main problem to establish the overarching objective, (2) the specific description and definitions of the current sub-problem, (3) historical code generated in preceding steps. This design requires the model to perform incremental development based on the existing code, thereby deeply probing its capacity to maintain logical coherence and interface consistency over long-horizon tasks.

\textbf{Prompts of LLM-Judge:} To evaluate the small number of questions that cannot be easily assessed using numerical results, we introduce an LLM-Judge mechanism using prompts designed by experts. These prompts leverage the comprehension capabilities of LLMs to guide them to act as senior code reviewers comparing the generated hypothesis with the reference ground truth. The prompts conduct a dual-dimensional evaluation encompassing functional equivalence, where LLM verifies if the generated code implements the same underlying business logic as the reference while disregarding superficial stylistic variations. Additionally, for syntactic structure, the prompts directs the judge to scrutinize control flow organization, modularity, and adherence to specific industrial coding standards. By requiring LLM to generate a step-by-step reasoning process before delivering a judgement, we ensure the evaluation rigorously captures both functional correctness and structural integrity.

\subsection{Evaluated Models}
To map the performance landscape of industrial code generation, we evaluate a comprehensive series of models ranging from proprietary SOTA flagships to diverse open-weights baselines. Our selection includes:
\begin{itemize}
    \vspace{-0.2cm}
    \item  \textbf{Proprietary Flagships:} The complete Claude 4.5 suite (Opus~\cite{claude45opus}, Sonnet~\cite{claude45sonnet}, Haiku~\cite{claude45haiku}), alongside the latest iterations of the GPT-5 (5.1, 5.2)~\cite{openaigpt52025} and Gemini (3-pro, 2.5-pro) families.
    \vspace{-0.2cm}
    \item \textbf{Open-Weights Leaders:} High-performance open models including Qwen3 (Max and the code-specialized Coder-Plus)~\cite{yang2025qwen3} and the MoE-based DeepSeek-V3.2-Exp~\cite{liu2025deepseek}.
    \vspace{-0.2cm}
    \item \textbf{Reasoning-Models:} To assess the impact of explicit reasoning strategies, we included both standard and thinking variants of Doubao-seed (1.6/1.8)~\cite{seedseed1}, GLM-4.7, and Kimi k2~\cite{team2025kimi}.
    \vspace{-0.2cm}
\end{itemize}

\subsection{Main Results}

We conducted a comprehensive evaluation of all models using IndustryCode, with Table 2 detailing the Pass@1 accuracy and overall scores across four industrial programming languages. At the foundational sub-problem level, overall model performance varied quite significantly, ranging from 27.7\% to 68.1\%. Claude 4.5 Opus demonstrated outstanding overall capability, consistently ranking in the top position in both sub-problems (68.1\%) and main problems (42.5\%). Claude 4.5 Sonnet and Gemini-3-pro follow closely, effectively establishing a solid baseline for SOTA performance. It is worth noting that various open-weights models exhibited remarkable competitiveness in standard industrial languages. In C++ sub-problems, Qwen3-Max achieves a high accuracy of 70.4\%, surpassing top proprietary models such as GPT-5.2 (67.9\%) and Gemini-3-pro (66.0\%). This clearly demonstrates the rapid progress of open models in mastering programming paradigms.

Regarding the evaluation of the main problems, the observed performance trends are consistent with those of the sub-problems, as shown in Figure 4. Claude 4.5 Opus maintained excellent capability across the majority of system-level settings, consistently outperforming other baselines with an overall accuracy of 42.5\%.


\begin{figure}[t]
  \vskip 0.2in
  \begin{center}
    \centerline{\includegraphics[width=\columnwidth]{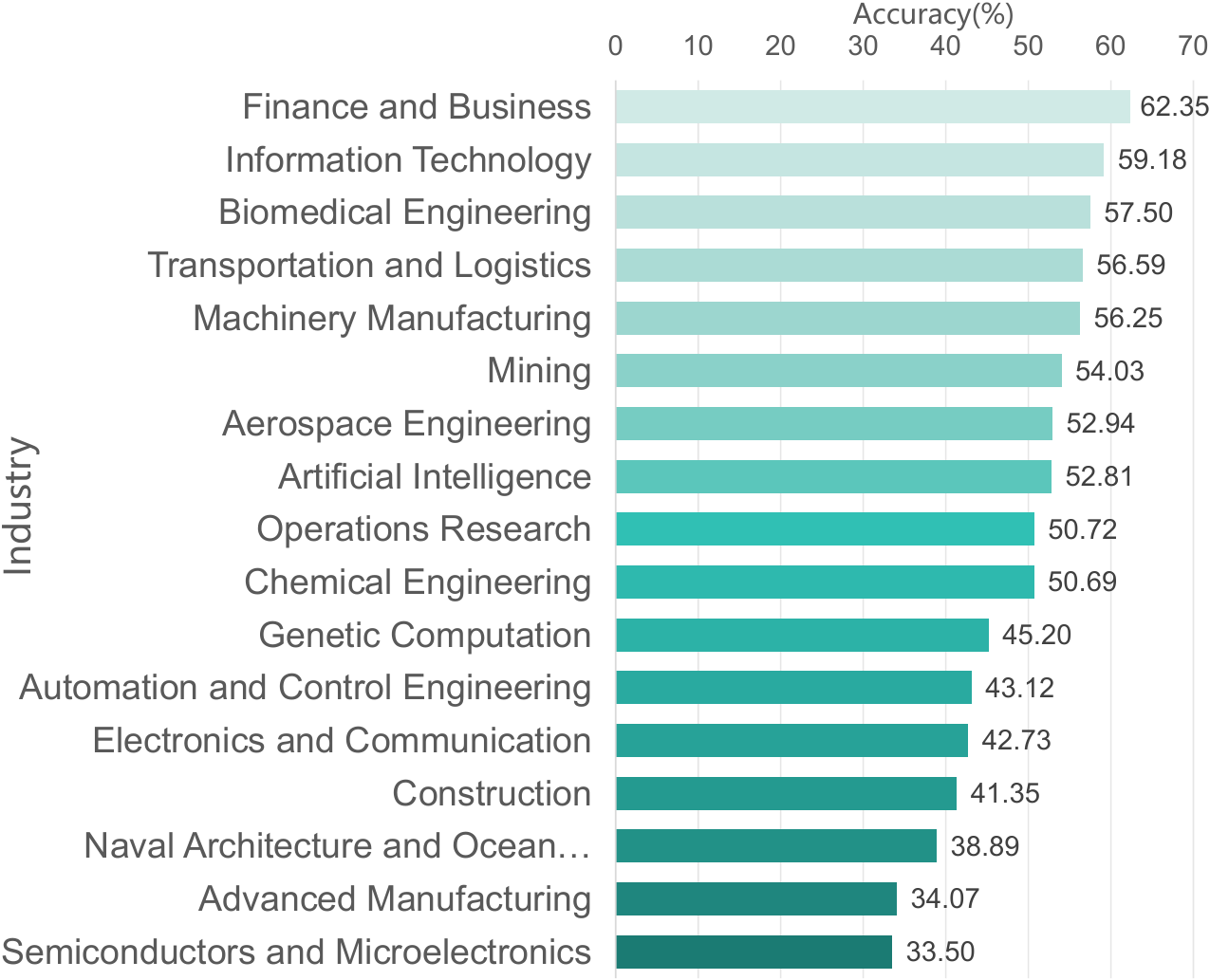}}
    \caption{
      \textbf{Performance distribution of sub-problem in IndustryCode.} The bar chart displays the average Pass@1 accuracy for each specific domain.
    }
    \vspace{-1cm}
    \label{icml-historical}
  \end{center}
\end{figure}
\subsection{Error Analysis and Failure Modes}
To identify the specific bottlenecks in industrial code generation, we conducted a fine-grained error analysis, as shown in Figure 5. The detailed distribution shows that model failures are mainly caused by syntactic deficits and semantic misalignment rather than logical errors. Syntax Error emerges as the leading failure mode (32.8\%), followed closely by Misunderstanding the Question (30.2\%), indicating that models struggle to adhere to the strict grammatical constraints of specialized industrial languages and fail to accurately convert complex engineering requirements into actionable logic. Furthermore, hallucination accounts for 19.6\% of errors, reflecting that when facing gaps in domain knowledge, the models tend to fabricate non-existent APIs. In contrast, Reasoning failure is relatively rare (8.4\%), indicating that current models have strong reasoning abilities but actually lack the specific vertical-domain corpora necessary for high-fidelity industrial applications.

\begin{figure}[t]
  \vskip 0.2in
  \begin{center}
    \centerline{\includegraphics[width=\columnwidth]{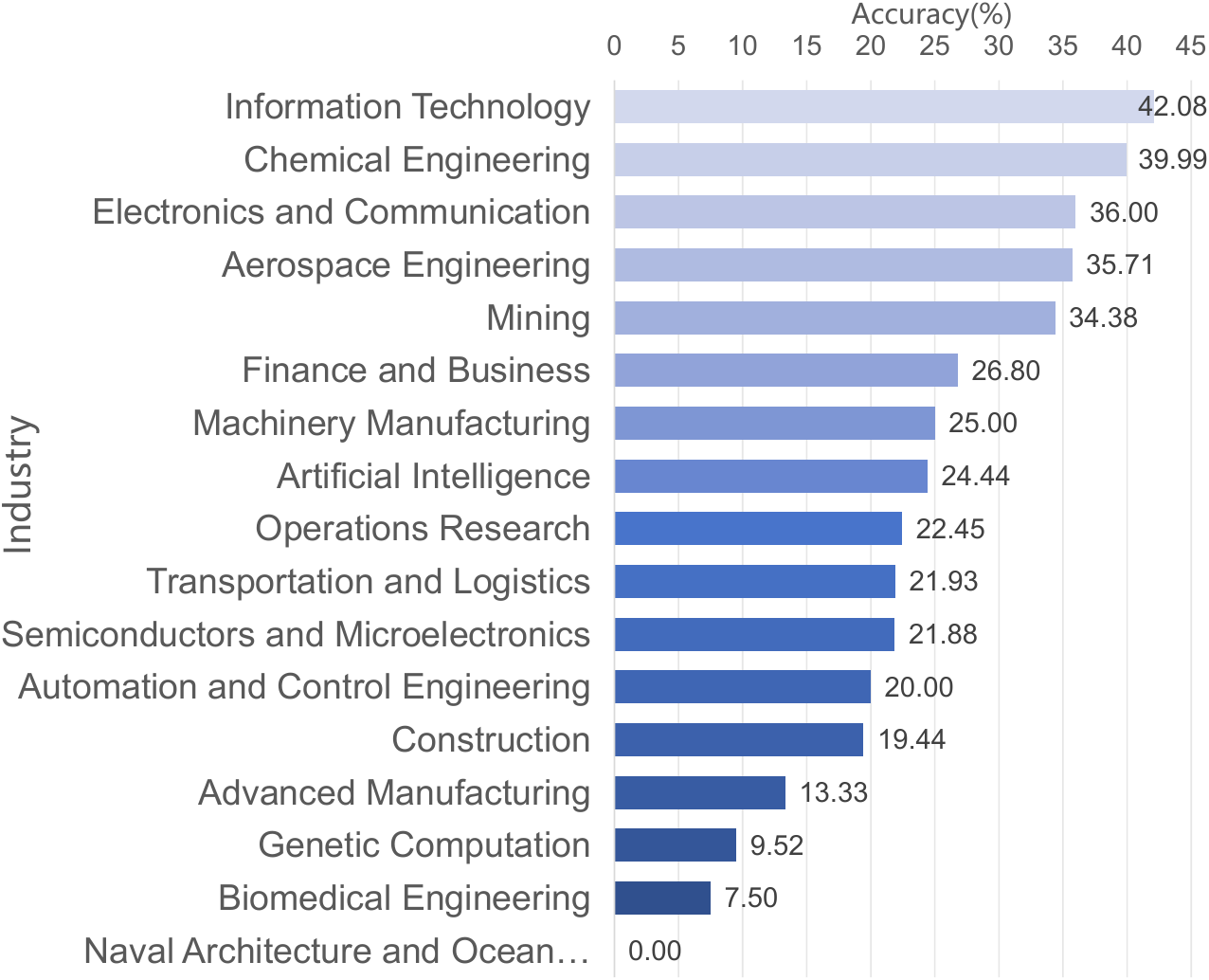}}
    \caption{
      \textbf{Performance distribution of mainproblem in IndustryCode.} The bar chart displays the average Pass@1 accuracy for each specific domain.
    }\vspace{-1cm}
    \label{icml-historical}
  \end{center}
\end{figure}

\subsection{Impact of Explicit Reasoning(Thinking Mode)}

To systematically assess the effectiveness of explicit reasoning strategies in industrial code generation, we conduct extensive ablation studies using representative  models, such as Doubao and GLM. We focus on evaluating performance shifts before and after activating the thinking mode. As illustrated in Figure 6, the results reveal the following findings:

\textbf{Improving performance through reasoning:} Empirical data indicates that enabling the thinking mode improves Pass@1 accuracy for certain large models, with average gains of approximately 4.70\% on sub-problems and 7.65\% on main problems for these benefited models. We attribute this gain to the Chain-of-Thought mechanism, which effectively decomposes complex mathematical derivations and logical constraints. This ability enables more accurate handling of complex numerical calculations and physical properties in industrial scenarios.

\textbf{Reasoning-Induced Errors:} However, it is worth noting that we observe that enabling the explicit thinking mode occasionally leads to a decline in the performance of some models. As shown in the error analysis in Figure 7, this performance decline is primarily caused by complexity-induced fragility and context confusion. The prolonged reasoning process often causes models to over-engineer solutions, generating unnecessary and complex logic, thereby significantly increasing the rates of syntax errors and hallucinations. Additionally, we observe a sharp rise in context confusion, where models fail to distinguish between their internal reasoning trace and the final code output, resulting in critical formatting violations.

\subsection{Domain-Specific Performance Analysis}
To delineate the boundaries of code generation correctness across diverse industrial domains, we analyzed the average success rates across 15 sub-domains. The results from Figure 3 (sub-problems) and Figure 4 (main problems) reveal a distinct capability divergence across different domains. In mathematically intensive and algorithmic domains, such as Finance and Business (62.35\%) and Information Technology (59.18\%), models achieve the highest accuracy. Notably, Information Technology demonstrates the strongest robustness in system-level integration, maintaining a 42.08\% accuracy on complex main problems. We attribute this success to the algorithmic certainty in these domains, enabling LLMs to leverage their inherent mathematical capabilities to derive solutions with high accuracy. 

Conversely, in highly specialized engineering fields such as Semiconductors and Microelectronics, the accuracy of sub-problems is limited to 33.50\%, while the accuracy of main problems is 21.88\%. Generating correct code in these domains requires specialized industrial knowledge and proficiency in specific languages. This creates a severe gap in domain knowledge, resulting in poor model performance.

\section{Related Works}
\textbf{Large Language Models for Code}
The evolution of large language models for code (Code LLMs) has shifted from fine-tuning general-purpose models towards 
specialised architecture design and training paradigms.  \cite{chen2021evaluating} pioneered this field by demonstrating 
the efficacy of large-scale code pre-training through Codex. Subsequently, open-source models advanced rapidly: Code 
Llama \cite{roziere2023code} and StarCoder \cite{li2023starcoder} significantly improved generation quality through 
long-context fine-tuning and the Fill-in-the-Model (FIM) task, respectively. In 2024, the DeepSeek-Coder  utilised 
MoE and multi-head latent attention (MLA) techniques \cite{zhu2024deepseek} to substantially optimise inference efficiency 
while maintaining SOTA performance. Regarding instruction-based approaches, 
WizardCoder \cite{luo2023wizardcoder} employs Evol-Instruct for automated generation of complex tasks, while Qwen 2.5 Coder \cite{hui2024qwen2} extends reasoning boundaries through massive high-quality corpora. Current research trends are shifting from static generation towards interactive reasoning and agents. OpenAI's o1 model introduced inference-time scaling, while Light et al. (2025) proposed SFS \cite{lightsfs} to optimise the generation-search process. Works such as the MCP protocol and DACO \cite{wu2024daco} focus on breaking down interaction barriers between models and real-world execution environments.

\textbf{Code Generation Benchmarks:} 
The evaluation of code generation has evolved from function-level synthesis standardized by HumanEval \cite{chen2021evaluating} and 
MBPP \cite{austin2021program} toward more complex reasoning tasks in APPS \cite{hendrycks2021measuring} and CodeContests \cite{li2022competition}. 
Recognizing the need for project-level interdependencies, recent benchmarks have pivoted to repository-scale contexts, with RepoBench \cite{liu2023repobench} and CrossCodeEval \cite{ding2023crosscodeeval} incorporating cross-file awareness, and SWE-bench \cite{jimenez2024swebench} introducing real-world issue resolution. Parallel efforts address evaluation nuances, such as scientific execution semantics in SciCode \cite{tian2024scicoderesearchcodingbenchmark} and data contamination in LiveCodeBench \cite{jain2024livecodebench}. While MultiPL-E \cite{cassano2023multiple} and HumanEval-X \cite{zheng2023codegeex} attempt to extend evaluation to multiple languages via translation, these approaches often result in non-idiomatic code and fail to capture the rigorous build environments and domain-specific paradigms in industrial workflows—a critical gap our work addresses by evaluating genuine polyglot professional capabilities.

\textbf{Industrial Capabilities Benchmark:} 
Beyond general-purpose programming, the assessment of industrial capabilities focuses on an LLM’s proficiency in specialized, high-value domains where code serves as a medium for solving complex problems. Recent vertical benchmarks have targeted hardware description languages for chip design, such as VerilogEval \cite{liu2023verilogeval} and RTLLM \cite{lu2023rtllm}, while others evaluate domain-specific pipelines in bioinformatics BioCoder \cite{tang2023biocoder} and end-to-end data science DSBench \cite{liu2024dsbench}. These efforts underscore that industrial competence requires deep semantic understanding and the mastery of complex toolchains, including precise API invocation Gorilla \cite{patil2023gorilla}, ToolBench \cite{qin2023tool} and advanced scientific reasoning SciBench \cite{wang2023scibench}. However, existing benchmarks are often confined to generic coding tasks, overlooking the specificity and diversity inherent in industrial applications. To address this, we evaluate LLMs across various industrial sectors and programming languages, aiming to rigorously assess their actual capabilities in real-world industrial code generation.

\section{Conclusion}
In this paper, we introduce IndustryCode, a comprehensive benchmark designed to evaluate real-world industrial code generation. 
IndustryCode encompasses 125 main problems and 579 sub-problems across diverse sectors such as finance, 
automation, and aerospace, covering programming languages including Python, C++, Matlab, and Stata. 
Evaluations conducted with multiple state-of-the-art AI models demonstrate the feasibility of our benchmark. 
We believe that IndustryCode will serve as a valuable guideline for developing future code language models tailored to various industrial applications.


\bibliography{icml_ref}
\bibliographystyle{icml2026}

\newpage
\appendix
\onecolumn


\section{Problem Descriptions and Examples}
\label{app:problem_examples}

\subsection{Library Dependency and Workload Characteristics}


\begin{figure}[H]
    \centering
    \begin{subfigure}[b]{\columnwidth}
        \centering
        \includegraphics[width=0.8\linewidth]{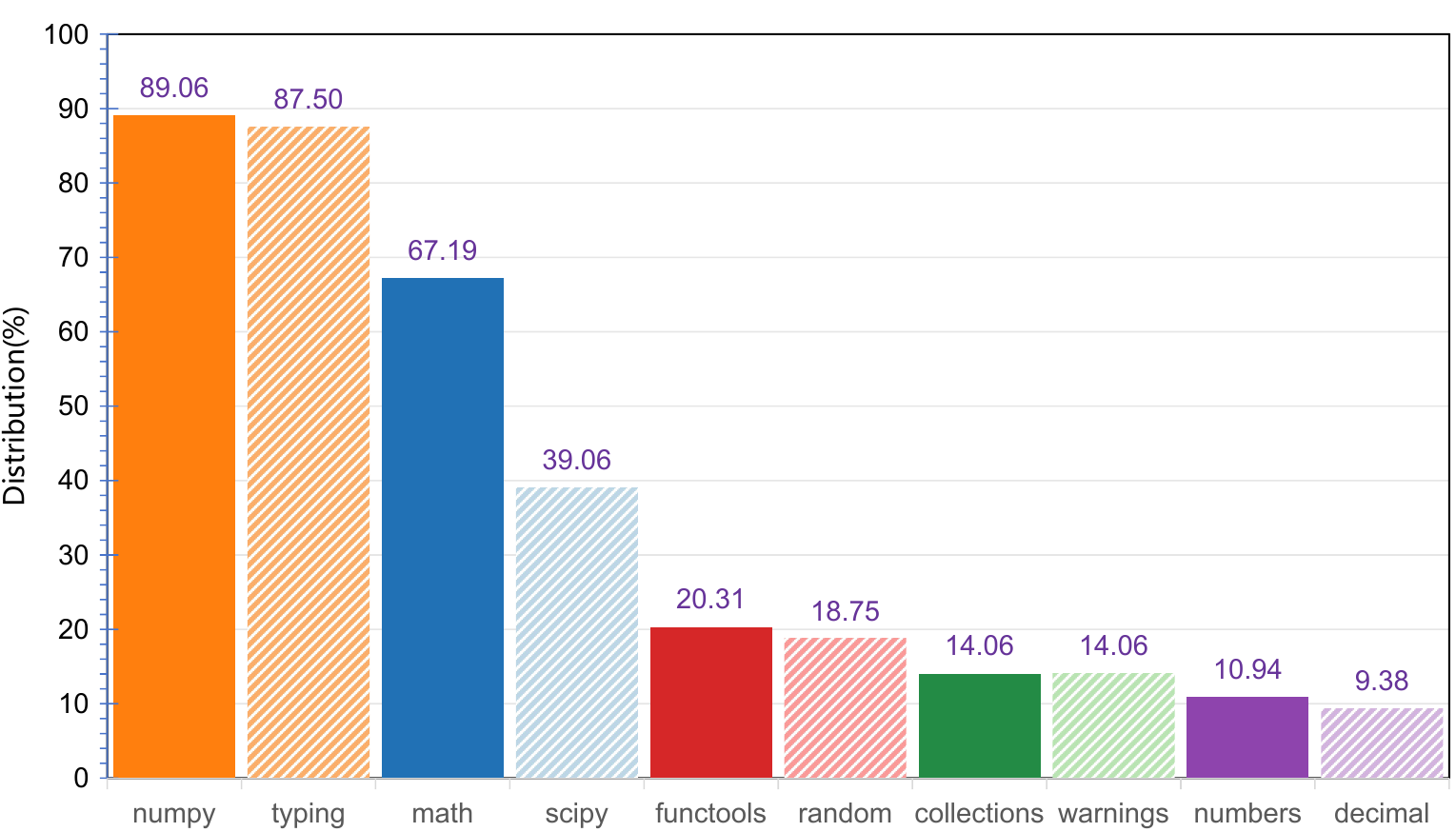} 
        \caption{}
        \label{fig:py_libs_a}
    \end{subfigure}
    
    \vspace{0.5cm} 

    \begin{subfigure}[b]{\columnwidth}
        \centering
        \includegraphics[width=0.8\linewidth]{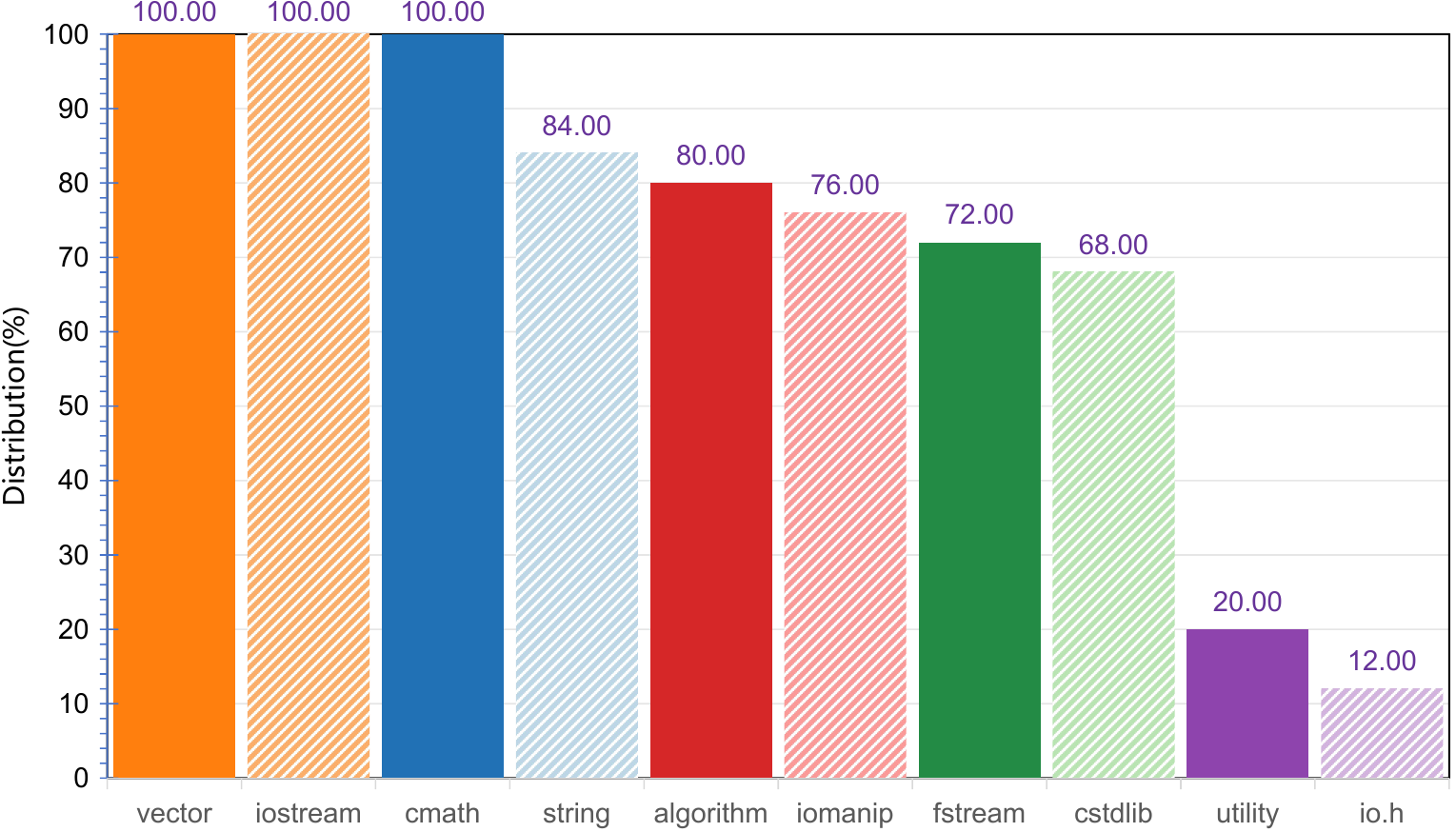} 
        \caption{}
        \label{fig:py_libs_b}
    \end{subfigure}

    \caption{\textbf{Dependencies of the Main Problem.} (a) Distribution of Python libraries used in IndustryCode. (b) Distribution of C++ header dependencies used in IndustryCode}
    \label{fig:py_libs}
\end{figure}



\subsection{Python-based LTI System Analysis}
\subsubsection*{1. System Overview and Library Dependencies}
\begin{figure}[H]
    \centering
    \includegraphics[width=0.95\linewidth]{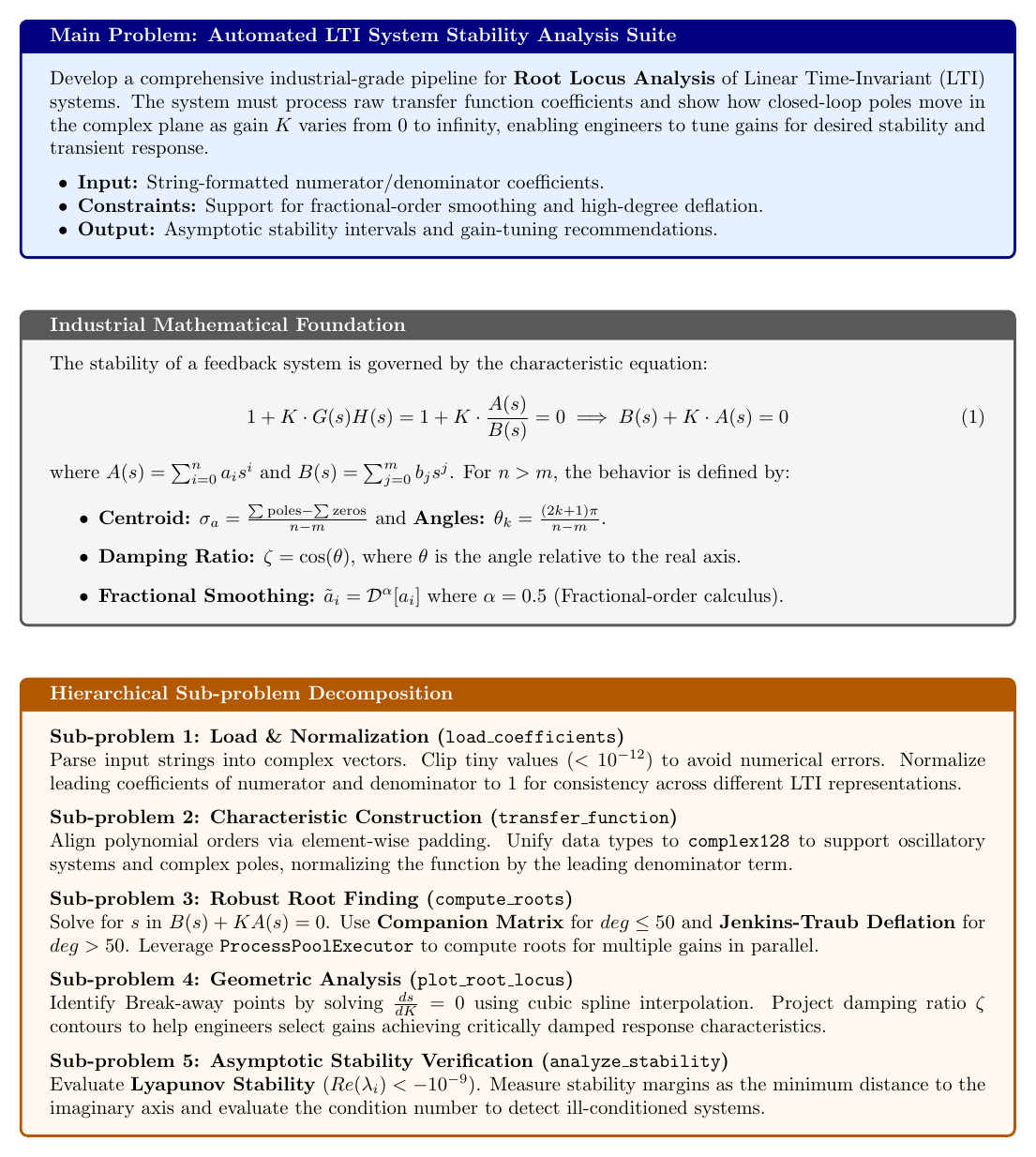}
    \caption{\textbf{Main Problem Description:} Overview of the Automated LTI System Stability Analysis Suite.}
    \label{fig:py_main}
\end{figure}

\subsubsection*{2. Core Algorithm Specifications}

\section*{Technical Details of Example Problems}

\begin{figure}[H]
    \centering
    \includegraphics[width=1\linewidth]{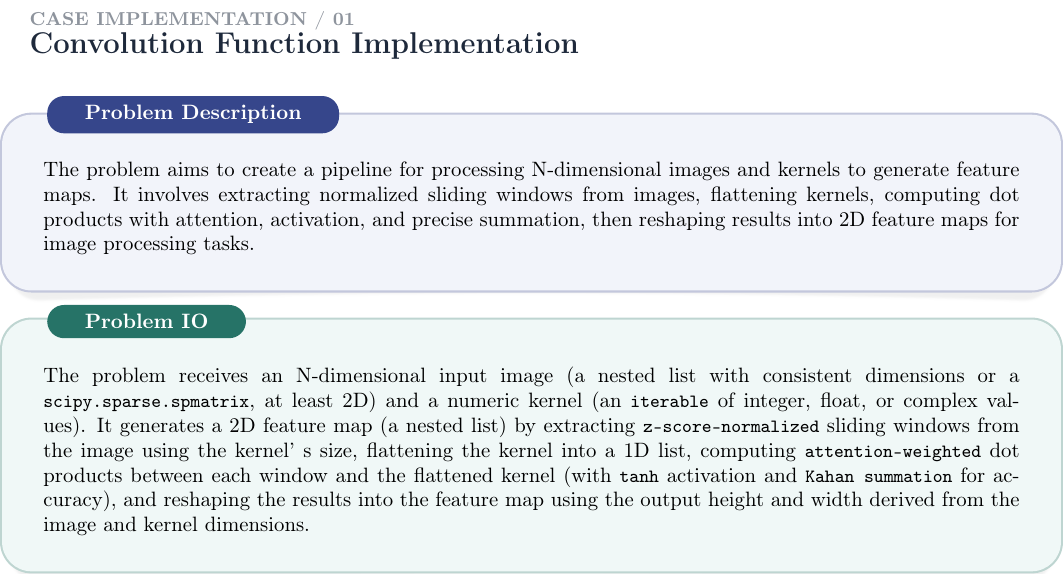}
    \caption{\textbf{Convolution Pipeline:} Specifications for N-dimensional image processing.}
    \label{fig:py_conv}
\end{figure}

The N-dimensional convolution pipeline implementation, as specified in Figure 11, is designed to evaluate the model's capacity for Algorithmic logic density and Architectural consistency within the IndustryCode framework. Unlike standard convolution problems that rely on black-box library calls such as \texttt{scipy.signal.convolve2d}, this benchmark requires the model to synthesize a multi-stage transformation from first principles. The process begins with a robust input interface that must reconcile dense nested structures with high-dimensional \texttt{scipy.sparse.spmatrix} objects. This requirement is representative of industrial data ingestion, where large-scale system matrices—often encountered in structural dynamics or Finite Element Analysis (FEA)—exhibit high degrees of sparsity that must be handled without excessive memory overhead.

A defining characteristic of this problem is the mandate for local \textit{z-score normalization} performed at the sliding window level. By transforming each extracted image patch to a zero-mean and unit-variance distribution before the dot-product operation, the algorithm forces the model to implement local contrast enhancement logic, which is critical for mitigating amplitude bias in sensor-based image data. Furthermore, the core computation is augmented with an \textit{attention-weighted mechanism}. This adds a layer of non-linear complexity to the standard multiply-accumulate operation, testing the model's ability to handle dynamic coefficient weighting—a prerequisite for implementing modern deep learning-based diagnostic frameworks within industrial environments.

To address the Numerical Precision demand highlighted in IndustryCode philosophy, the specification necessitates the use of \textbf{Kahan Summation} for the accumulation of dot products. In high-frequency signal processing and iterative mechanical solvers, the accumulation of millions of floating-point residuals often leads to catastrophic cancellation and significant precision loss. The Kahan algorithm provides a compensated summation method that tracks the lost low-order bits through a running correction term, ensuring that the final feature maps remain physically accurate. Finally, the reconstruction phase evaluates the model's proficiency in topological mapping, requiring the 1D results to be correctly indexed and reshaped into a 2D manifold based on derived spatial dimensions. This synthesis of modular logic, numerical stability, and coordinate mapping serves as a rigorous stress test for the gap of domain knowledge  identified in our evaluation.

\begin{figure}[H]
    \centering
    \includegraphics[width=1.0\linewidth]{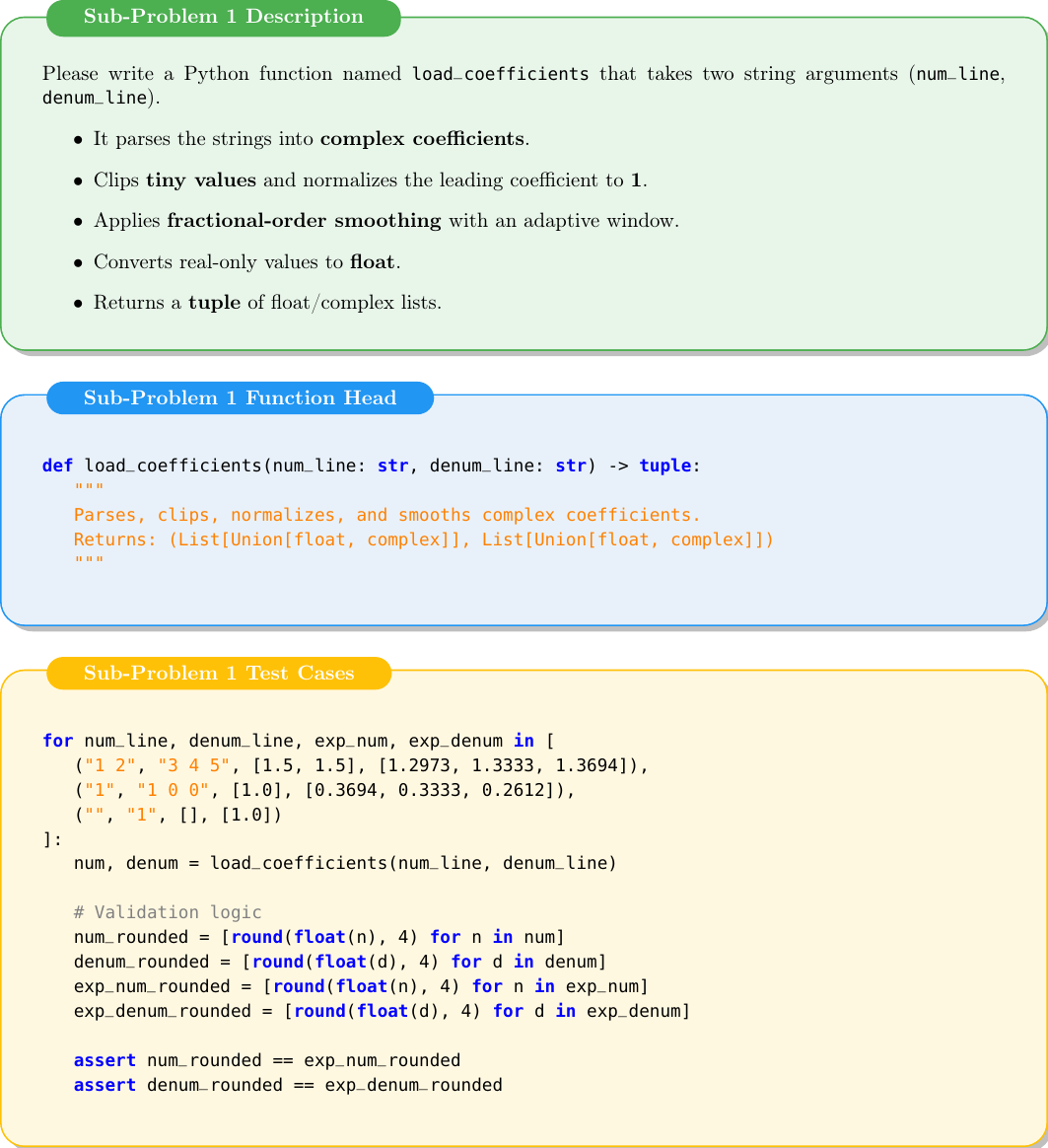}
    \caption{\textbf{Load Coefficients Utility:} Detailed sub-problem requirements for coefficient normalization.}
    \label{fig:py_transfer}
\end{figure}

\begin{figure}[H]
    \centering
    \includegraphics[width=1.0\linewidth]{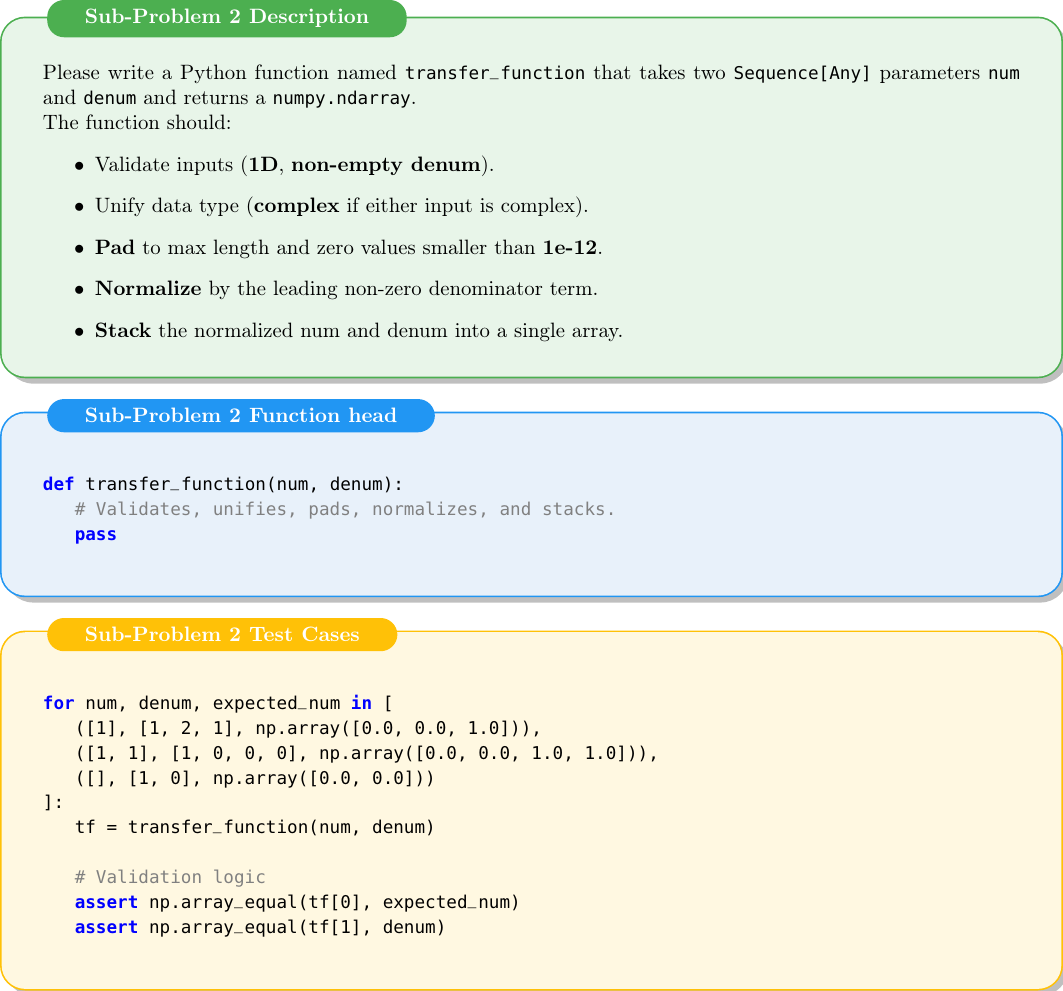}
    \caption{\textbf{Transfer Function Utility:} Detailed sub-problem requirements for transfer function.}
    \label{fig:py_transfer}
\end{figure}


\begin{figure}[H]
    \centering
    \includegraphics[width=1.0\linewidth]{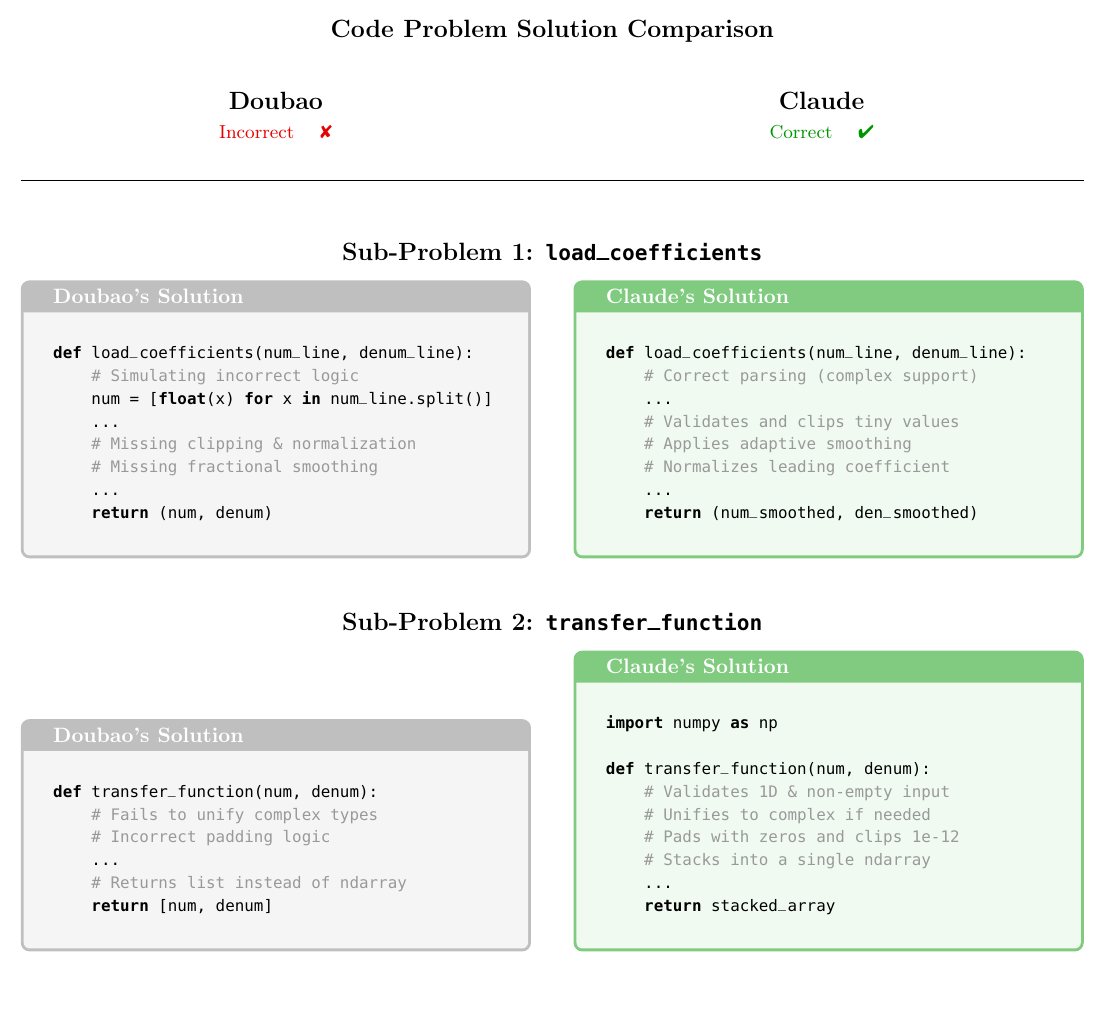}
    \caption{\textbf{Models comparison:} Doubao and Claude as Examples}
    \label{fig:py_transfer}
\end{figure}

Figure 14 illustrates the distinct logical gap between the models. Doubao's solutions tend to be oversimplified, failing to address critical technical constraints such as data type unification and rigorous input validation. In contrast, Claude demonstrates a comprehensive grasp of the requirements, correctly implementing complex processing logic to ensure the output strictly adheres to the specified standards.

\subsection{MATLAB-based Optimization and Control}

\subsubsection*{1. Mathematical Definitions}
\begin{figure}[H]
    \centering
    \includegraphics[width=1\linewidth]{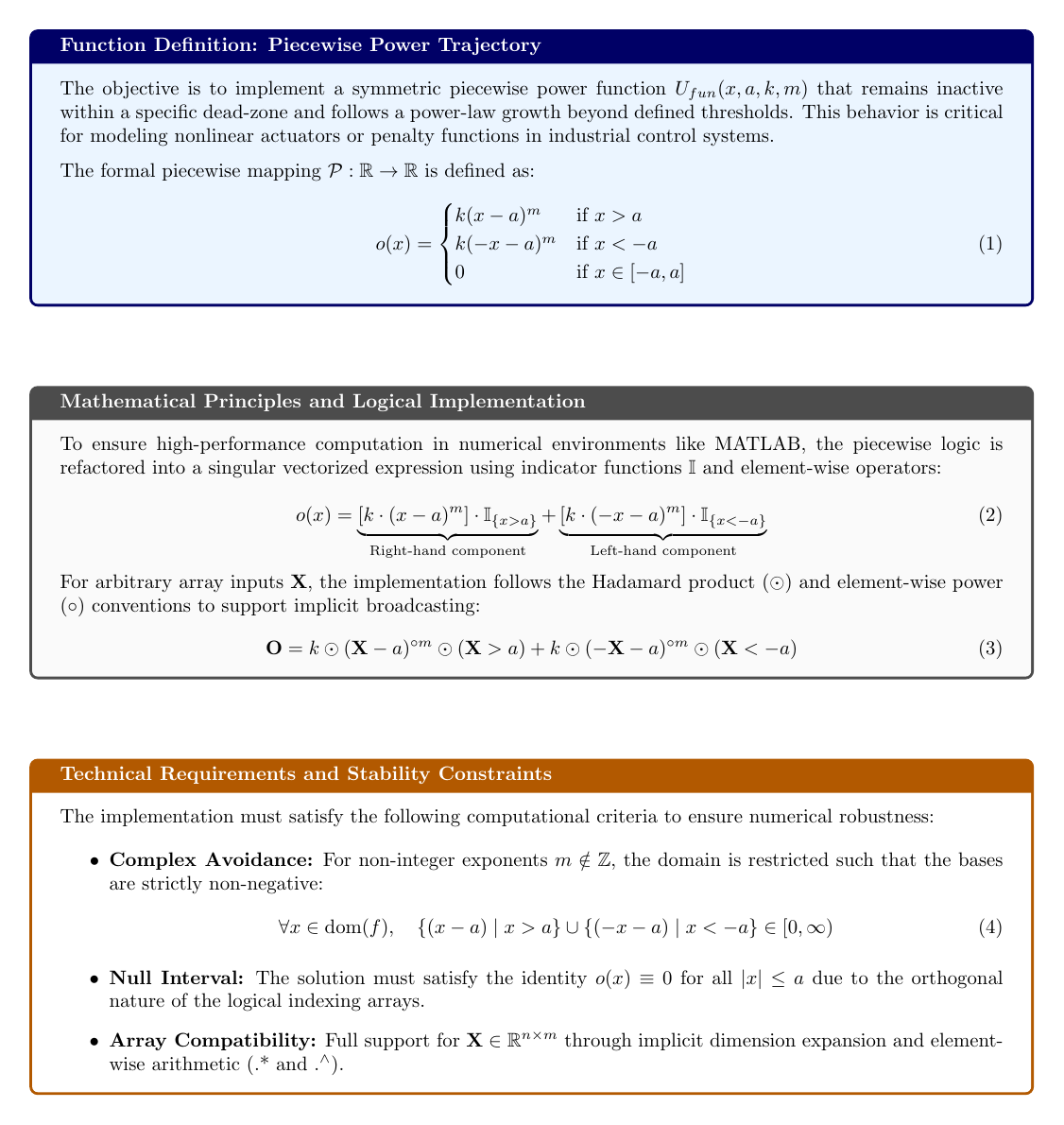}
    \caption{\textbf{Piecewise Power Trajectory:} Mathematical definition and vectorized logic for the control function.}
    \label{fig:mat_math}
\end{figure}
\raggedbottom 
\subsubsection*{2. Optimization Strategy (GWO)}
\begin{figure}[H]
    \centering
    \includegraphics[width=1.0\linewidth]{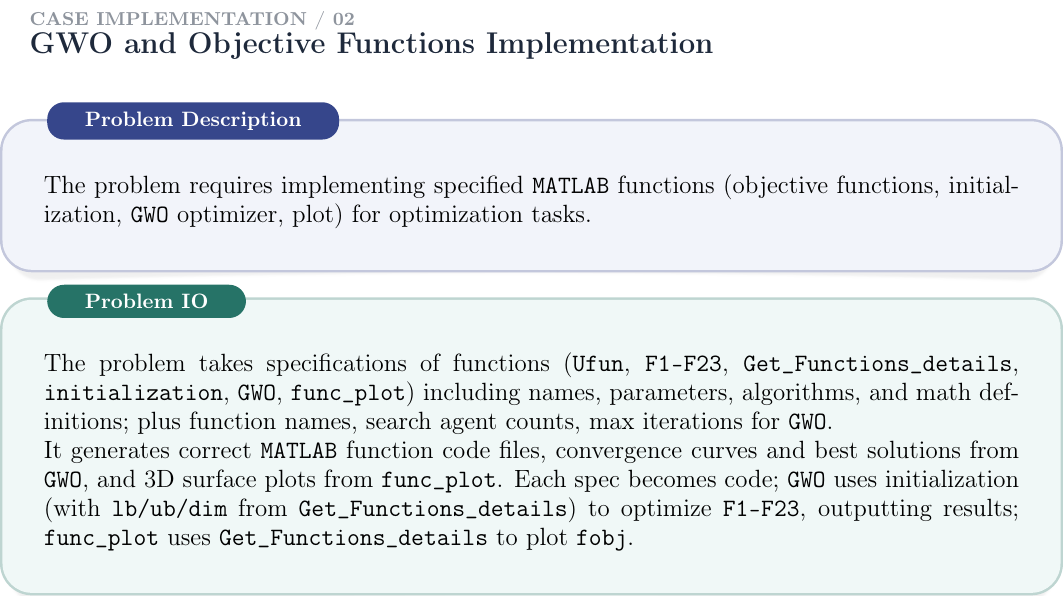}
    \caption{\textbf{GWO Implementation:} Problem description for the Grey Wolf Optimizer problem.}
    \label{fig:mat_gwo}
\end{figure}


In accordance with IndustryCode design philosophy, the MATLAB-based optimization suite is implemented through a hierarchical factorization of complex engineering workflows. The high-level optimization layer, exemplified by the Gray Wolf Optimizer (GWO) in Figure 16, evaluates the model's capacity to manage stochastic search processes and multi-agent state updates within a scientific computing context. This problem goes beyond simple script generation, requiring the Large Language Model (LLM) to synthesize a modular architecture that separates the core heuristic logic from boundary constraint management (\textit{lb, ub}) and problem-specific metadata. Such a structure is representative of professional-grade industrial software, where maintainability and the ability to swap objective functions without re-engineering the optimizer are paramount.

A critical challenge identified in this benchmark is the Domain Knowledge Gap regarding MATLAB's vectorized semantics. Effective optimization in industrial scenarios necessitates the use of implicit broadcasting and element-wise matrix operations to handle high-dimensional search spaces efficiently. The coupling between the GWO optimizer and its fitness evaluation relies on the precision of underlying piecewise trajectories. To maintain numerical stability across millions of iterations, the sub-components must be architected with rigorous logical indexing, ensuring that discontinuities in the power functions do not lead to non-physical gradients or divergent solutions. The following sub-problem, \texttt{Ufun}, provides the granular algorithmic requirements necessary to construct these robust fitness landscapes, serving as a fundamental unit of the broader control pipeline.


\subsubsection*{3. Detailed Function Requirements}
\begin{figure}[H]
    \centering
    \includegraphics[width=1.0\linewidth]{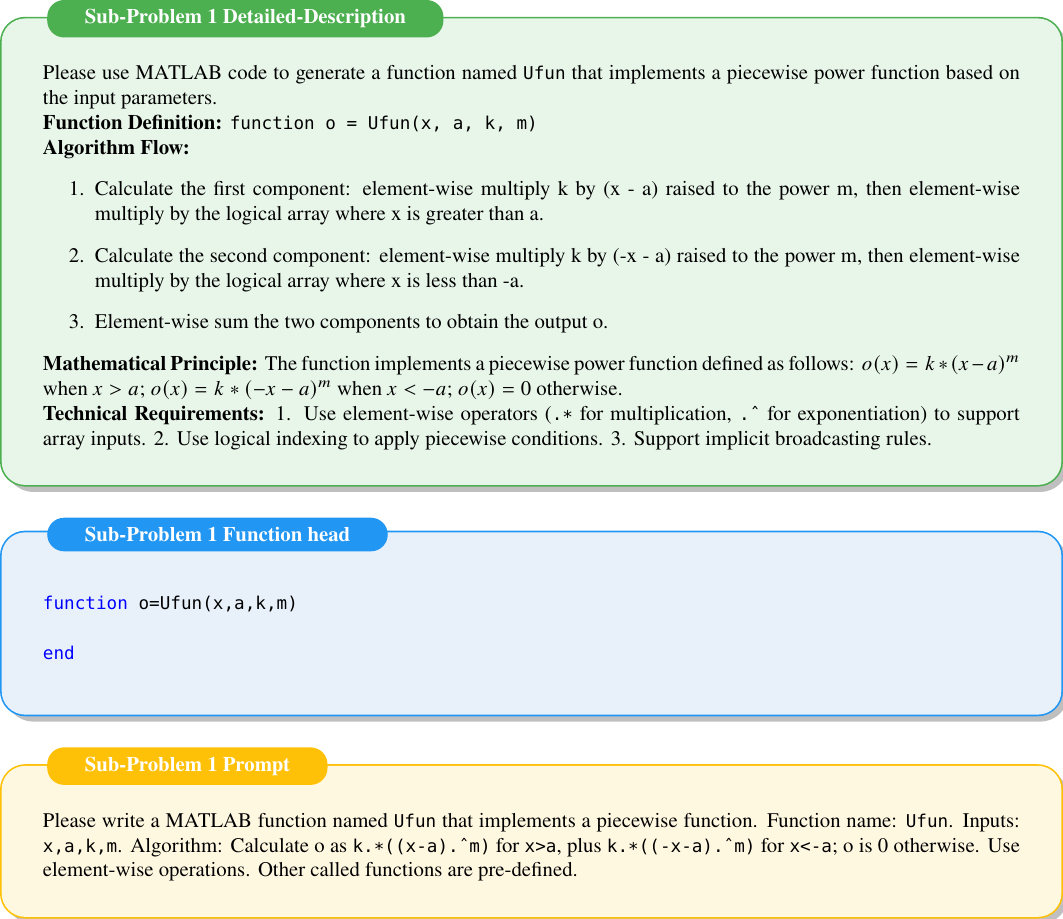}
    \caption{\textbf{Ufun Implementation:} Step-by-step algorithm flow for the piecewise power function.}
    \label{fig:mat_ufun}
\end{figure}

\clearpage

\section{Comprehensive Model Performance Analysis}
\label{app:model_performance}

\subsection{Cross-Industry Performance Evaluation}


\stepcounter{figure}
\begin{figure}[H]
    \centering
    \includegraphics[width=1.0\linewidth]{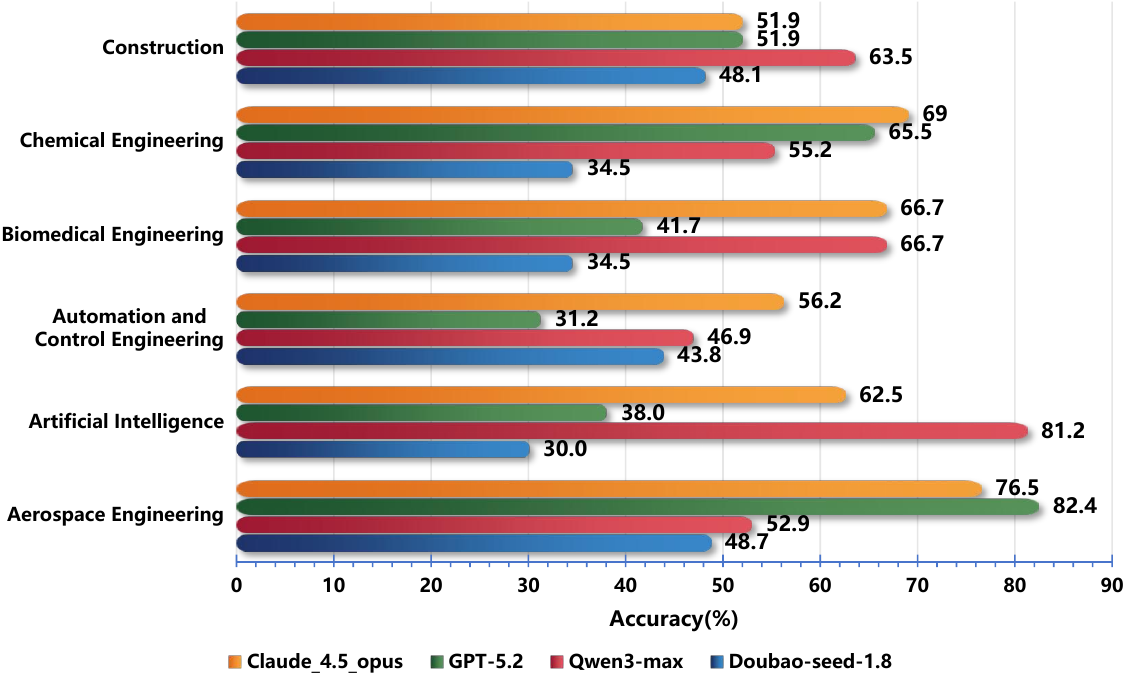}
    \\[1em] 

    \includegraphics[width=1.0\linewidth]{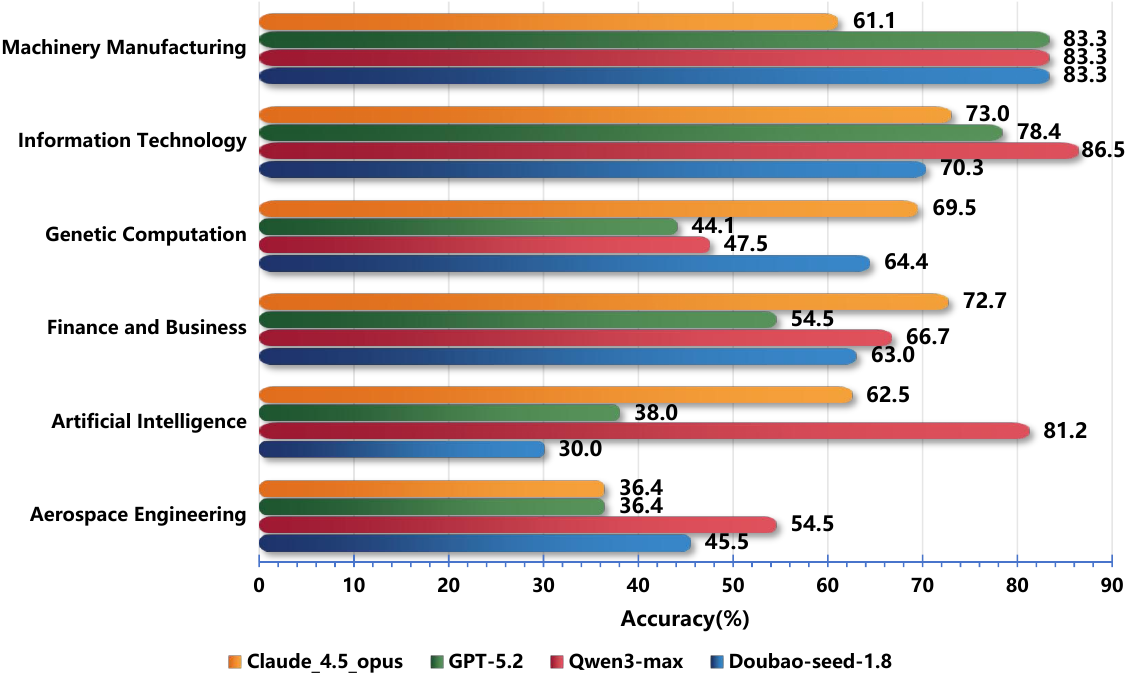}
    \\[1em]
    \label{fig:ind_bar}
\end{figure}

\begin{figure}[t!] 
    \centering
    \ContinuedFloat 
    \includegraphics[width=0.9\linewidth]{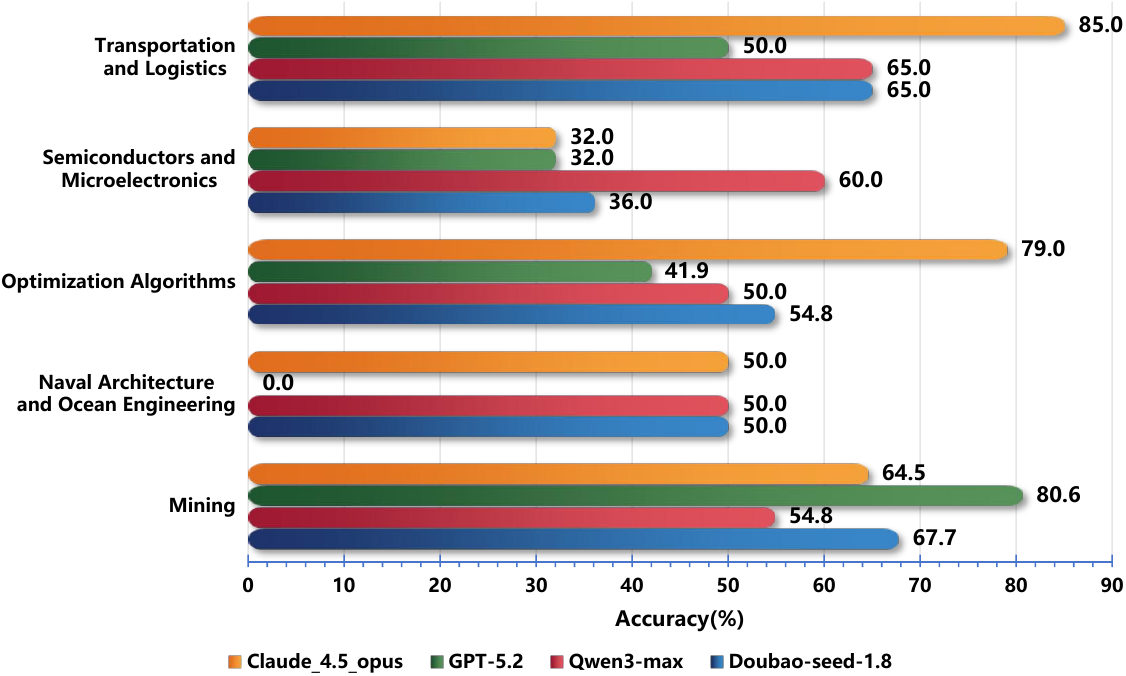}
    
    \caption{\textbf{Industry Success Rates:} Comparative performance across 16 domains.}
    \label{fig:ind_bar}
\end{figure}

Figure 18 shows the success rates of two representative open-source models and two closed-source models across different industries. The results indicate that while closed-source models like Claude 4.5 Opus and GPT-5.2 lead in overall scores, open-source models are highly competitive in specific sectors. In certain tasks, Qwen even outperformed Claude 4.5 Opus, which is currently considered the strongest model overall.

Specifically, in the fields of AI and Information Technology, Qwen3-max achieved accuracy rates of $81.2\%$ and $86.5\%$, both significantly higher than those of Claude 4.5 Opus. Most notably, in the specialized field of Semiconductors and Microelectronics, Qwen3-max beat the closed-source models with a score of $60\%$. This demonstrates that by optimizing for vertical industries, open-source models can not only narrow the gap with top-tier closed-source models but even overtake them in key areas. This offers a more cost-effective solution for industry-specific deployment.


\subsection*{B.1.1 Cross-Industry Performance Overview}

Figure 18 presents a comparative analysis of four state-of-the-art models (Claude 4.5 Sonnet, GPT-5.2, Doubao-seed-1.8, and Qwen3-Max) across 16 industrial domains. The selection of these specific models is strategic: they serve as prominent representatives of proprietary and open-weights architectures from the world's two dominant AI ecosystems (US and China). This comparison examines the effects of distinct training data distributions and architectural philosophies—specifically dense versus Mixture-of-Experts (MoE) architectures—on industrial adaptability. The results indicate a substantial performance disparity between digital software domains and physical engineering domains. Sectors rooted in pure software, such as Information Technology and Artificial Intelligence, consistently demonstrate high pass rates, with Qwen3-Max peaking at 86.5\%. In contrast, hardware-centric disciplines like \textit{Semiconductors} and \textit{Naval Architecture} exhibit significantly lower accuracy and higher inter-model variance, suggesting that current LLM architectures face challenges in generalizing from software logic to physically constrained engineering problems.


\subsection*{B.1.2 Deep Dive: Performance Disparities and Architectural Causes}

\textbf{1. Architectural Misalignment in Semiconductors and Microelectronics}

Performance in the \textit{Semiconductors and Microelectronics} sector is consistently suppressed across all evaluated models, with GPT-5.2 recording approximately 32.0\%, while Qwen3-Max and Claude achieve roughly 60\%. The fundamental cause appears to be the paradigm divergence between the sequential nature of LLM token generation and the concurrent nature of Hardware Description Languages (HDLs) such as Verilog. Models trained primarily on sequential software languages (e.g., Python or C++) frequently misapply blocking assignments in contexts requiring non-blocking assignments, leading to timing violations in circuit design problems.

\textbf{2. Knowledge Retrieval in Physical Engineering}

In heavy engineering fields such as \textit{Aerospace Engineering} (82.4\%) and \textit{Mining} (80.6\%), GPT-5.2 demonstrates superior performance. This suggests that the pre-training corpus for GPT-5.2 likely contains a higher density of technical specifications, standard operating procedures (SOPs), and industry standards (e.g. NASA specifications or geological survey reports). Success in these domains relies less on abstract reasoning and more on the precise knowledge retrieval of domain-specific literature that other models may have filtered out during data curation.

\textbf{3. Mathematical Reasoning versus Pattern Matching}

A significant performance divergence is observed in \textit{Optimization Algorithms}. Claude 4.5 Sonnet (79.0\%) outperforms Qwen3-Max ($\sim$50\%) and GPT-5.2 ($\sim$42\%) by a wide margin. Optimization problems necessitate multi-step logical deduction rather than pattern recognition. The architecture of Claude, which is optimized for long-context reasoning, enables the construction of valid mathematical proofs and the maintenance of complex algorithmic constraints. Conversely, other models often rely on heuristic approximation based on similar training examples, resulting in lower accuracy for abstract mathematical problems.

\textbf{4. Mixture-of-Experts Efficiency in IT and Manufacturing}

Qwen3-Max achieves the highest individual score in the benchmark (86.5\% in IT) and leads in \textit{Machinery Manufacturing}. These results support the efficacy of the Mixture-of-Experts (MoE) architecture in standardized coding problems. By routing code-generation problems to specialized parameter groups trained on extensive repositories of Python, C++, and Matlab, the MoE architecture maximizes parameter efficiency and mitigates the interference often observed in dense models when processing disparate domain requirements.


\subsection*{B.1.3 Summary Statistics Table}

The following table summarizes the top-performing models across key sectors, illustrating the shift in optimal architecture from software-centric MoE models to those specialized in abstract reasoning or retrieval-heavy engineering problems.

\begin{table}[h!]
\centering
\caption{Dominant Models by Industrial Domain (Pass@1 Accuracy)}
\label{tab:industry_performance}
\begin{tabular}{@{}llcl@{}}
\toprule
\textbf{Industry Domain} & \textbf{Leading Model} & \textbf{Score (\%)} & \textbf{Primary Performance Factor} \\ \midrule
Information Technology & Qwen3-Max & \textbf{86.5} & MoE Expert Routing \\
Transportation \& Logistics & Claude 4.5 Sonnet & 85.0 & Contextual Understanding \\
Machinery Manufacturing & Qwen3-Max & 83.3 & Pattern Matching (G-code/PLC) \\
Aerospace Engineering & GPT-5.2 & 82.4 & Technical Knowledge Retrieval \\
Artificial Intelligence & Claude 4.5 Sonnet & 81.2 & Algorithm Implementation \\
Mining & GPT-5.2 & 80.6 & Specification Adherence \\
Optimization Algorithms & Claude 4.5 Sonnet & 79.0 & Chain-of-Thought Reasoning \\
Semiconductors & Claude / Qwen3-Max & 60.0 & \textit{(Concurrent Logic Limitations)} \\
\bottomrule
\end{tabular}
\end{table}

\subsection{problem-Specific Performance Metrics}

\begin{figure}[H]
    \centering
    \includegraphics[width=0.95\linewidth, height=0.80\textheight, keepaspectratio]{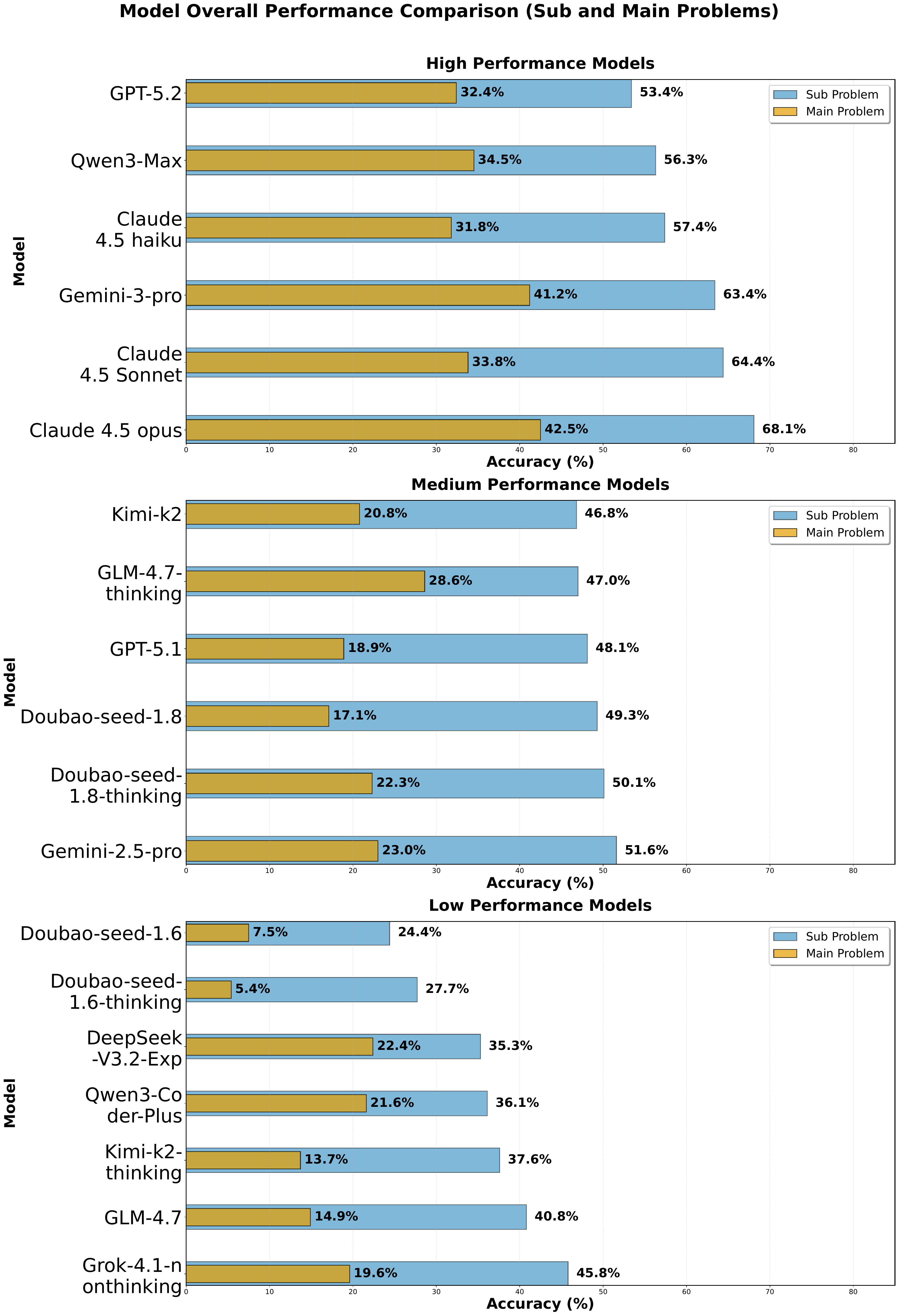}
    \caption{\textbf{Tiered Model Breakdown:} Pass@1 Accuracy for High, Medium, and Low performance groups. Note the consistent gap between Main problem (purple) and Sub problem (blue) results.}
    \label{fig:stacked_bar}
\end{figure}

Figures 19 present the performance of various Large Language Models (LLMs) on autonomous problems. The evaluation encompasses two dimensions: Main Problem (representing the completion of the ultimate objective) and Sub Problem (representing the successful execution of individual intermediate steps within the process). The following section provides a detailed analysis of these results.

\subsubsection*{B.2.1. Tiered Performance Analysis}

\vspace{0.5em}
\noindent  Model performance on this benchmark exhibits a distinct stratified distribution. Figure 19 highlights the disparity between state-of-the-art models and mid-range or trailing models. 
\textbf{Leading Models:} Claude 4.5 Opus ranks first with a significant margin, demonstrating superior overall alignment. It is followed by Claude 4.5 Sonnet and Gemini-3-pro, constituting the top tier.
\textbf{High-Performance Segment:} The first tier is primarily occupied by models from Anthropic (Claude series), Google (Gemini series), Alibaba (Qwen series), and OpenAI (GPT series).
\textbf{Performance Disparity:} There is a substantial magnitude of difference in performance between the top-scoring Claude 4.5 Opus (Main Problem 42.5\%) and models in the medium performance group, such as Kimi-k2 (20.8\%).

\subsubsection*{B.2.2 Tiered Breakdown Analysis.}

\vspace{0.5em}
\noindent \textbf Figure 19 classifies models into three specific performance tiers, facilitating an analysis of technical maturity across different architectures.

\noindent \textbf{High Performance Models.} 
Members: GPT-5.2, Qwen3-Max, Claude 4.5 (Haiku, Sonnet, Opus), Gemini-3-pro. 
Characteristics: The Main Problem success rate for this group generally ranges between 30\% -- 42.5\%, with Sub Problem success rates between 53\% -- 68\%. 
Key Observations: Claude 4.5 Opus achieves the highest scores on both Main and Sub Problems. Gemini-3-pro demonstrates exceptional performance on Sub Problem completion (63.4\%), surpassed only by the high-end Claude models.

\noindent \textbf{Medium Performance Models.} 
Members: Kimi-k2, GLM-4.7, GPT-5.1, Doubao-seed-1.8 (and its reasoning-enhanced version), Gemini-2.5-pro. 
Characteristics: Main Problem success rates decline significantly to the 17\% -- 28\% range. 
Observation: Reasoning-enhanced variants (denoted as thinking) occupy significant positions in this tier. For instance, GLM-4.7-thinking reached 28.6\%, outperforming most other models in the same group.

\noindent \textbf{Low Performance Models.} 
Members: Doubao-seed-1.6 series, DeepSeek-V3.2-Exp, Qwen3-Coder-Plus, Grok-4.1, etc. 
Characteristics: Main Problem success rates are generally below 20\%, with some models (e.g., Doubao-seed-1.6) achieving single-digit accuracy. This suggests these models have not yet met the minimum threshold for complex autonomous problem execution.

\subsubsection*{B.2.3 Core Insight: The Execution Gap.}
\vspace{0.5em}
\noindent \textbf A critical observation from the evaluation is the Execution Gap: models consistently score significantly higher on Sub Problems than on Main Problems.
\textbf{Phenomenon:} Sub Problem accuracy consistently exceeds Main Problem accuracy across all evaluated models.
\textbf{Root Cause Analysis:} This indicates that while models are capable of executing discrete, atomic actions (such as information retrieval or code generation), they struggle to orchestrate these sequential steps to achieve the final autonomous goal. This reflects primary limitations in current agents regarding multi-step planning, long-term context maintenance, and error recovery.
\textbf{Gap Quantification:} Even for the best-performing Claude 4.5 Opus, there remains a significant divergence of 25.6\% between Sub Problems (68.1\%) and Main Problems (42.5\%). This implies that despite high success rates in individual steps, sequential error accumulation often leads to overall problem failure.

\subsubsection*{B.2.4 Core Insight: The Execution Gap.}
\vspace{0.5em}
\noindent \textbf{Claude Series (Anthropic):} Demonstrates consistent superiority. From the lightweight Haiku to the most powerful Opus, all members appear in the High Performance group, highlighting the architectural advantages of this family in agentic problems.
\textbf{Gemini Series (Google):} Shows clear generational advancement. Gemini-3-pro significantly outperforms the previous generation Gemini-2.5-pro (situated in the Medium group), demonstrating substantial capability improvements through model iteration.
\textbf{Reasoning-Enhanced Models:} The comparison within the medium tier confirms the efficacy of reinforced reasoning. For example, \textbf{Doubao-seed-1.8-thinking} (Main Problem 22.3\%) outperforms its standard counterpart \textbf{Doubao-seed-1.8} (Main Problem 17.1\%), indicating that integrating chain-of-thought or reasoning mechanisms effectively enhances autonomous problem-solving capabilities.

\subsection{Language-Specific Analysis}


\begin{figure}[htbp]
    \centering
    \includegraphics[width=1.0\linewidth]{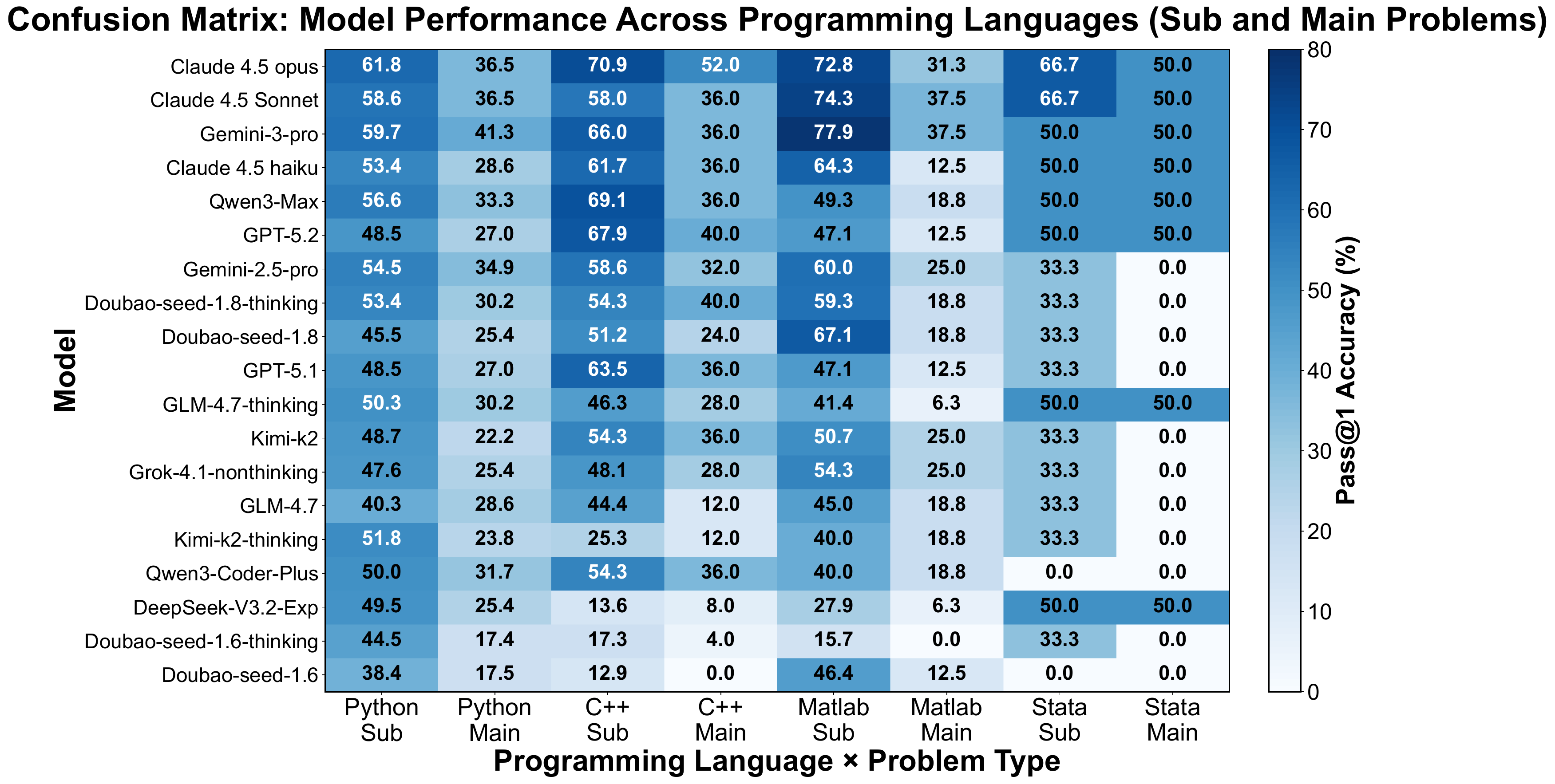} 
    \caption{\textbf{Confusion Matrix of Success Rates:} The heatmap highlights a significant performance gap between high-resource languages (Python/C++) and low-resource languages (Stata/MATLAB), as well as between Sub-problems and Main-problems.}
    \label{fig:confusion_matrix}
\end{figure}

\subsubsection{Quantitative Disparity Across Languages and Problems}
\label{subsubsec:quantitative_disparity}

The visualization in Figure~\ref{fig:confusion_matrix} reveals a structural stratification in model capabilities. 
First, there is a distinct stratification based on language popularity. Widely adopted languages such as \textit{Python} and \textit{C++} (columns 1-2) exhibit high saturation in accuracy, with top-tier models like Claude 4.5 Opus achieving over 60\% on sub-problems. In sharp contrast, domain-specific languages like \textit{Stata} (columns 7-8) show near-zero performance for the majority of models, resulting in a region of negligible performance in the heatmap.
Second, a significant degradation is observed when moving from atomic problems (Sub) to holistic problems (Main). For instance, while \textit{Python} sub-problems maintain high accuracy, the main problems see a performance drop of approximately 30-50\% across all models. This indicates that while current LLMs excel at generating code snippets, they struggle significantly with system-level engineering and complex logic flow.

\subsubsection{The Data Availability Disparity in Pre-training Data}
\label{subsubsec:data_disparity}

The fundamental cause of the performance dichotomy observed above can be attributed to the disparity in data availability within LLM training distributions. \textit{Python} and \textit{C++} benefit from massive open-source ecosystems (e.g., GitHub), providing models with billions of high-quality training tokens. Conversely, languages like \textit{Stata} and \textit{MATLAB} are largely proprietary and used within closed academic or industrial circles. The scarcity of public repositories for these languages results in models that lack proficiency in these domains. The heatmap confirms that model architecture alone (even in GPT-5 series or Claude 4.5) cannot overcome the fundamental lack of domain-specific training data.

\subsubsection{Syntactic Dependencies and Reasoning Limitations}
\label{subsubsec:reasoning_limitations}

Beyond data scarcity, the nature of the syntax and problem complexity plays a critical role. 
The sharp decline in main problems suggests that models rely heavily on \textbf{pattern matching} rather than \textbf{genuine long-context reasoning}. Sub-problems often resemble LeetCode-style algorithms commonly found in training sets, whereas main problems require managing global variables and multi-step logic.
Furthermore, specialized languages like \textit{Stata} possess \textbf{state-dependent syntax} that differs fundamentally from the logic-flow of \textit{C++} or \textit{Python}. Standard tokenizers, optimized for English and mainstream code, likely fragment these specialized commands into non-semantic sub-tokens, forcing the model to expend disproportionate attention resources to parse basic syntax, thereby degrading its reasoning capacity.

\newcommand{\SafeIncludeGraphics}[2][]{%
    \IfFileExists{#2}{%
        \includegraphics[#1]{#2}%
    }{%
        \centering
        \fbox{\begin{minipage}{0.9\linewidth}\centering
            \vspace{1cm}
            \textbf{Figure Source Not Found:}\\
            \texttt{#2}\\
            \vspace{0.5cm}
            \small(Please upload the corresponding PDF file to Overleaf)\\
            \small(File Name must match exactly)
            \vspace{1cm}
        \end{minipage}}%
    }%
}

\section{Model Code Comparison and Analysis}
\subsection{Theoretical Framework: Mechanisms of Generative Failure}
\label{sec:theoretical_framework}

The integration of Chain-of-Thought (CoT) reasoning into large language models introduces a fundamental paradox: while logical reasoning capabilities improve, low-level syntactic precision often degrades. We categorize the primary mechanisms driving this phenomenon into three distinct cognitive failures:

\begin{itemize}
    \item \textbf{Cognitive Over-correction (Reasoning-Induced Domain Mismatch):} The extended reasoning process leads the model to prioritize generalized engineering practices over specific domain constraints. The model applies broad programming conventions, such as strict type safety or fixed-width definitions, which are statistically prevalent in training data but unsuitable for the specific immediate context. This phenomenon results from the application of global standards to incompatible specific scopes rather than a failure of context retention, as observed in the incorrect substitution of integer types in legacy formats such as BMP.

    \item \textbf{Contextual Interference (Inhibition Failure):} Models act as both semantic processors and syntactic generators. A failure in inhibition control occurs when the semantic context of the problem description (e.g., theoretical formulas, background narratives) leaks into the syntactic generation stream, appearing as hallucinated variables or verbatim text insertion.
    
    \item \textbf{State Decay in Autoregressive Generation:} As the sequence length increases, the fidelity of the Key-Value (KV) cache diminishes regarding global state tracking. This leads to violations of the One Definition Rule (ODR), where the model fails to retain the information that a structure has already been defined and attempts to redefine it within the same scope.
\end{itemize}

\subsection{Doubao-1.6 Failure Modes and Mode Comparison}
\label{sec:doubao_analysis}

\subsubsection{Taxonomy of Structural Failures}

\begin{figure}[H]
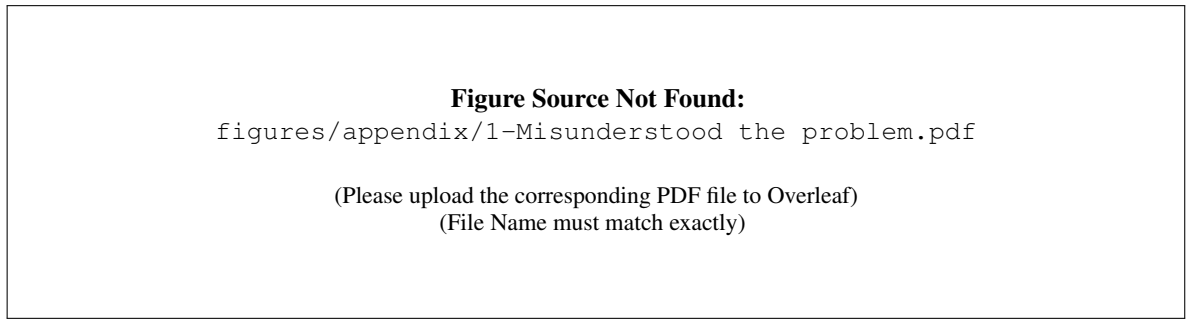

    \centering
    \SafeIncludeGraphics[width=0.95\textwidth]{figures/appendix/1-Misunderstood the problem.pdf}
    \caption{\textbf{Type Degradation and Header Incoherence.} The model fails to adhere to strict type constraints (e.g., \texttt{int}), reverting to other generic integer types, and includes irrelevant library headers.}
    \label{fig:doubao_error_1}
\end{figure}

\begin{figure}[H]
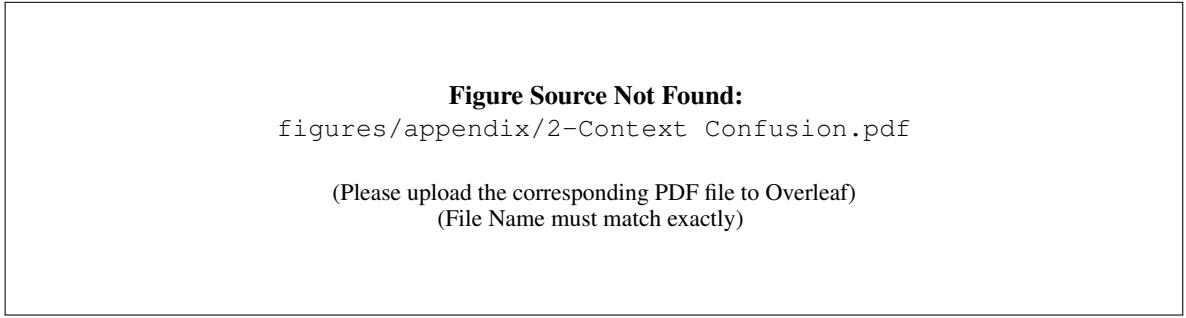

    \centering
    \SafeIncludeGraphics[width=0.95\textwidth]{figures/appendix/2-Context Confusion.pdf}
    \caption{\textbf{Contextual Confusion.} Semantic descriptions from the prompt (e.g., NDVI formulas) breach the code boundary, appearing as invalid syntax or undefined variables.}
    \label{fig:doubao_error_2}
\end{figure}

\begin{figure}[H]
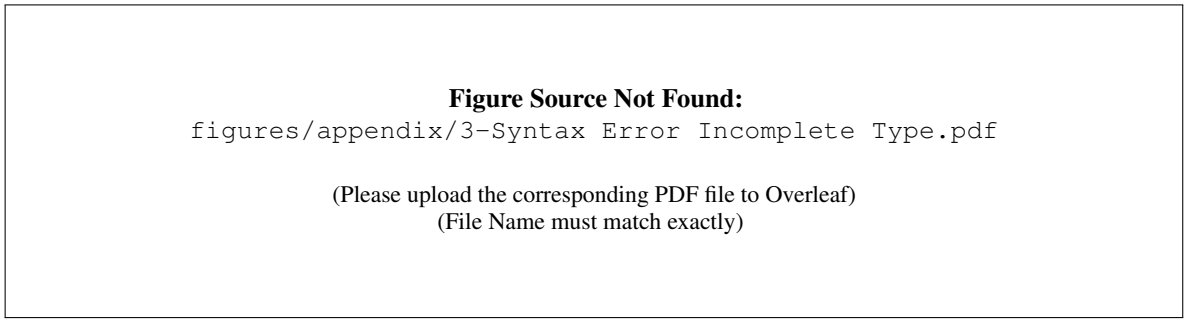

    \centering
    \SafeIncludeGraphics[width=0.95\textwidth]{figures/appendix/3-Syntax Error Incomplete Type.pdf}
    \caption{\textbf{Logic Truncation.} The generation terminates prematurely due to token budget exhaustion during the reasoning phase, leaving critical syntactic structures unclosed.}
    \label{fig:doubao_error_3}
\end{figure}

We have isolated four characteristic failure modes in Doubao-1.6 that persist despite the activation of reasoning modules. These failures are visualized in Figures \ref{fig:doubao_error_1} through \ref{fig:doubao_error_4}.

\noindent \textbf{Mechanistic Analysis:} The observed failure provides evidence of \textbf{Cognitive Over-correction}, where the extended reasoning process drives the prioritization of generalized modern programming practices over specific legacy constraints. In this instance, the model systematically replaces the standard signed \texttt{int} with \texttt{uint32\_t}, adhering to the heuristic that dimension descriptors should be unsigned and fixed-width. This substitution violates the BMP file specification, which explicitly requires signed integers for fields such as \texttt{biHeight} to determine bitmap orientation through positive or negative values. Consequently, the error stems from the active misapplication of generic type-safety rules to an incompatible specific domain, rather than from a passive degradation of context retention.

\noindent \textbf{Mechanistic Analysis:} This failure represents a critical mode collapse between the latent reasoning space and the syntactic generation space. Large Language Models utilize specific attention heads, often termed Induction Heads, to retrieve and copy relevant entities from the context. In this scenario, the Induction Heads become over-activated by the semantic embeddings of the problem description (e.g., Mahalanobis distance, K-Nearest Neighbors). The error stems from a failure in inhibition control. Ideally, the model should attend to the prompt for logic extraction but inhibit the verbatim copying of text when in the syntactic generation mode. Here, the boundary blurs: the model creates a latent scope where it hallucinates that variables mentioned in the prompt (like $K$) are globally accessible in the code. This violates strict scope resolution rules in C++, indicating that the internal semantic representation of the variable $K$ overpowered the syntactic necessity of declaring it as a formal function parameter.

\noindent \textbf{Mechanistic Analysis:} The structural truncation observed here is a direct consequence of token budget exhaustion and RLHF misalignment. The reasoning process is token-expensive; it consumes a significant portion of the maximum output length ($L_{max}$) of the model. As the generation approaches this hard limit, the model employs computational parsimony. Internal heuristics, likely tuned via Reinforcement Learning from Human Feedback (RLHF), penalize verbosity. To avoid a hard cutoff mid-sentence, the probability distribution for the End-of-Sequence (EOS) token rises artificially. This pressure forces the model into simplified generation modes—substituting complex, token-heavy logic with comments or placeholders. Crucially, this tendency toward premature completion often overrides the syntactic stack, causing the model to terminate generation before popping all open scopes (i.e., closing braces), resulting in code that is logically outlined but syntactically broken.

\begin{figure}[H]
    \centering
    \SafeIncludeGraphics[width=0.95\textwidth]{figures/appendix/4-Syntax Error Redefinition.pdf}
    \caption{\textbf{State Tracking Failure.} The model loses track of the global symbol table, resulting in the redundant redefinition of data structures within the same scope.}
    \label{fig:doubao_error_4}
\end{figure}

\noindent \textbf{Mechanistic Analysis:} This error illustrates the repetition curse in autoregressive generation, amplified by the stateless nature of the Transformer. The model relies entirely on the Key-Value (KV) Cache to track the state of the generated program. However, reasoning introduces a disconnect between planning and execution. When the model decomposes a complex problem, its reasoning trace outlines the sequence (e.g., define header, then define data). The model generates the header correctly in the first step. However, as it proceeds to the subsequent step, the attention mechanism may re-attend to the plan (which mentions defining the header) rather than the output (which already contains the definition). Lacking a symbolic symbol table, the model fails to recognize that the state \texttt{BMPInfoHeader} is previously defined. It interprets the semantic concept of the structure in its context window as a signal to generate it again, violating the One Definition Rule (ODR). This highlights a fundamental limitation in the ability of the model to maintain global state consistency across long reasoning horizons.

\subsubsection{Quantitative Analysis: The Cost of Reasoning}

Table \ref{tab:failure_modes_full} presents a comprehensive quantitative shift in failure modes when switching from No-Thinking to Thinking mode. The data reveals a distinct Pareto trade-off: the activation of the thinking process effectively minimizes logical errors, with \textbf{Reasoning Failure} dropping significantly from 11.8\% to 0.8\%. However, this improvement comes at the cost of stability across all other dimensions.

Specifically, we observe a systemic degradation in instruction adherence and generation precision:
\begin{itemize}
    \item \textbf{Context Confusion} exhibits the most dramatic surge, rising from 1.9\% to 29.2\%, indicating a reduced ability to strictly adhere to prompt constraints.
    \item \textbf{Problem Misinterpretation} doubles from 15.0\% to 30.4\%, while \textbf{Syntax Error} rates increase from 4.4\% to 9.4\%.
    \item Furthermore, the Thinking mode introduces higher rates of generative instability, with \textbf{Hallucination} increasing from 8.7\% to 12.4\% and \textbf{Truncation} errors rising slightly from 1.2\% to 1.9\%.
\end{itemize}

\begin{table}[htbp]
    \centering
    \caption{Performance Metrics: Doubao-1.6 Reasoning Mode vs. Standard Mode}

    \label{tab:failure_modes_full}
    \resizebox{0.9\linewidth}{!}{%
    \begin{tabular}{@{}lcccccc@{}}
        \toprule
        \textbf{Model Mode} & \textbf{\shortstack{Context\\Confusion}} & \textbf{\shortstack{Problem\\Misinterp.}} & \textbf{\shortstack{Syntax\\Error}} & \textbf{Hallucination} & \textbf{Truncation} & \textbf{\shortstack{Reasoning\\Failure}} \\
        \midrule
        No-Thinking & \textbf{1.9\%} & \textbf{15.0\%} & \textbf{4.4\%} & \textbf{8.7\%} & \textbf{1.2\%} & 11.8\% \\
        Thinking    & 29.2\% & 30.4\% & 9.4\% & 12.4\% & 1.9\% & \textbf{0.8\%} \\
        \bottomrule
    \end{tabular}
    }
    \vspace{0.1cm}
    \par
    \footnotesize{\textit{Note: Data derived from the distribution of failure modes (Figure 6). While logical reasoning improves significantly, the Thinking mode incurs penalties in adherence (Context Confusion) and faithfulness (Hallucination).}}
\end{table}

\noindent \textbf{Mechanistic Interpretation:} The divergence in error distribution—where logical performance improves (low Reasoning Failure) while adherence and precision degrade (high Context Confusion)—suggests a phenomenon of \textit{Reasoning-Induced Heuristic Bias}.

In the reasoning mode, the model generates extensive intermediate traces to decompose complex logic. While this process enhances algorithmic problem-solving, it concurrently alters the generative context by activating generalized engineering priors. This introduces two critical failure mechanisms:

\begin{itemize}
    \item \textbf{Cognitive Over-correction (Heuristic Prioritization):} The reasoning process reinforces global best practices found in the training corpus, such as modern type safety or strict memory alignment. Consequently, the model actively prioritizes these generalized heuristics over the specific constraints of the prompt. This results in the misapplication of valid but contextually inappropriate engineering standards to incompatible local domains, as evidenced by the substitution of signed types with unsigned variants in legacy file formats.
    
    \item \textbf{Contextual Interference (State Persistence):} The extended natural language reasoning phase creates a persistent semantic state that interferes with the transition to strict syntactic generation. The model fails to fully suppress the explanatory context when switching to code output. This persistence leads to the erroneous inclusion of non-executable descriptions or hallucinated headers alongside valid code blocks, as the semantic embeddings of the problem description remain active during the syntactic generation phase.
\end{itemize}

\subsection{Comparative Mechanistic Analysis of Generative Failures}
\label{sec:comparative_analysis}

This section contrasts the performance of Doubao-1.6 against Claude-4.5 Opus across the four identified failure modes, providing a deep mechanistic dissection of the divergence in model performance.

\subsubsection{Case 1: Misunderstood the problem}
\begin{figure}[H]
    \centering
    \SafeIncludeGraphics[width=0.95\textwidth]{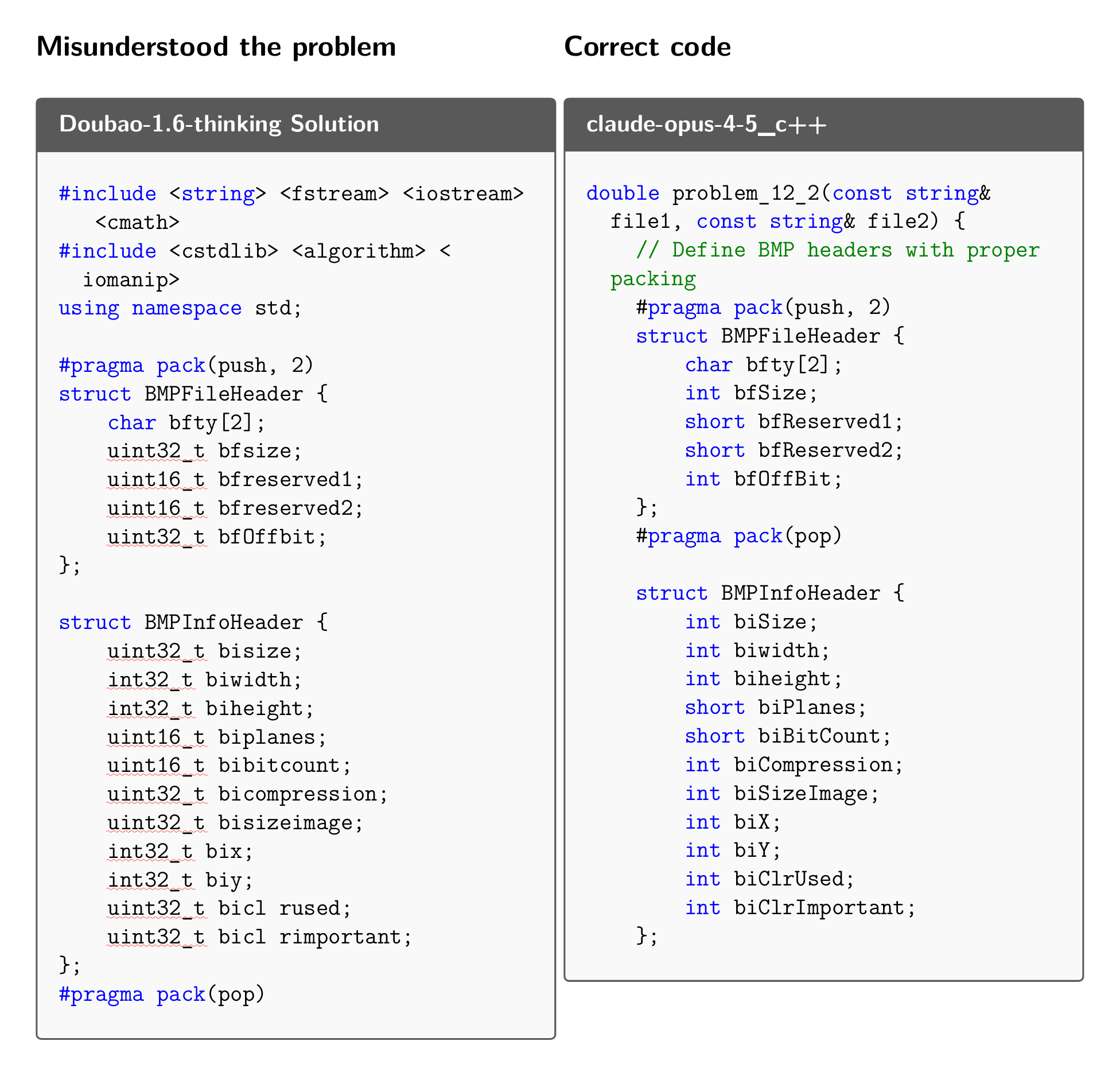}
    \caption{\textbf{Comparative Analysis of Type Fidelity.} Left: Doubao (Degraded). Right: Claude (Strict).}
    \label{fig:compare_1}
\end{figure}

\textbf{Analysis of Claude's Performance:}
In the header definition task, Claude strictly adheres to the legacy specifications of the BMP format. The model correctly identifies that the \texttt{biHeight} field requires a signed integer type (implemented here as standard \texttt{int}) to support the orientation logic of the protocol, where negative values indicate a top-down bitmap. This approach demonstrates a prioritization of specific domain constraints over generalized engineering heuristics, avoiding the erroneous substitution of fixed-width unsigned types observed in the comparison model.

\subsubsection{Case 2: Context Confusion}
\begin{figure}[H]
    \centering
    \SafeIncludeGraphics[width=0.95\textwidth]{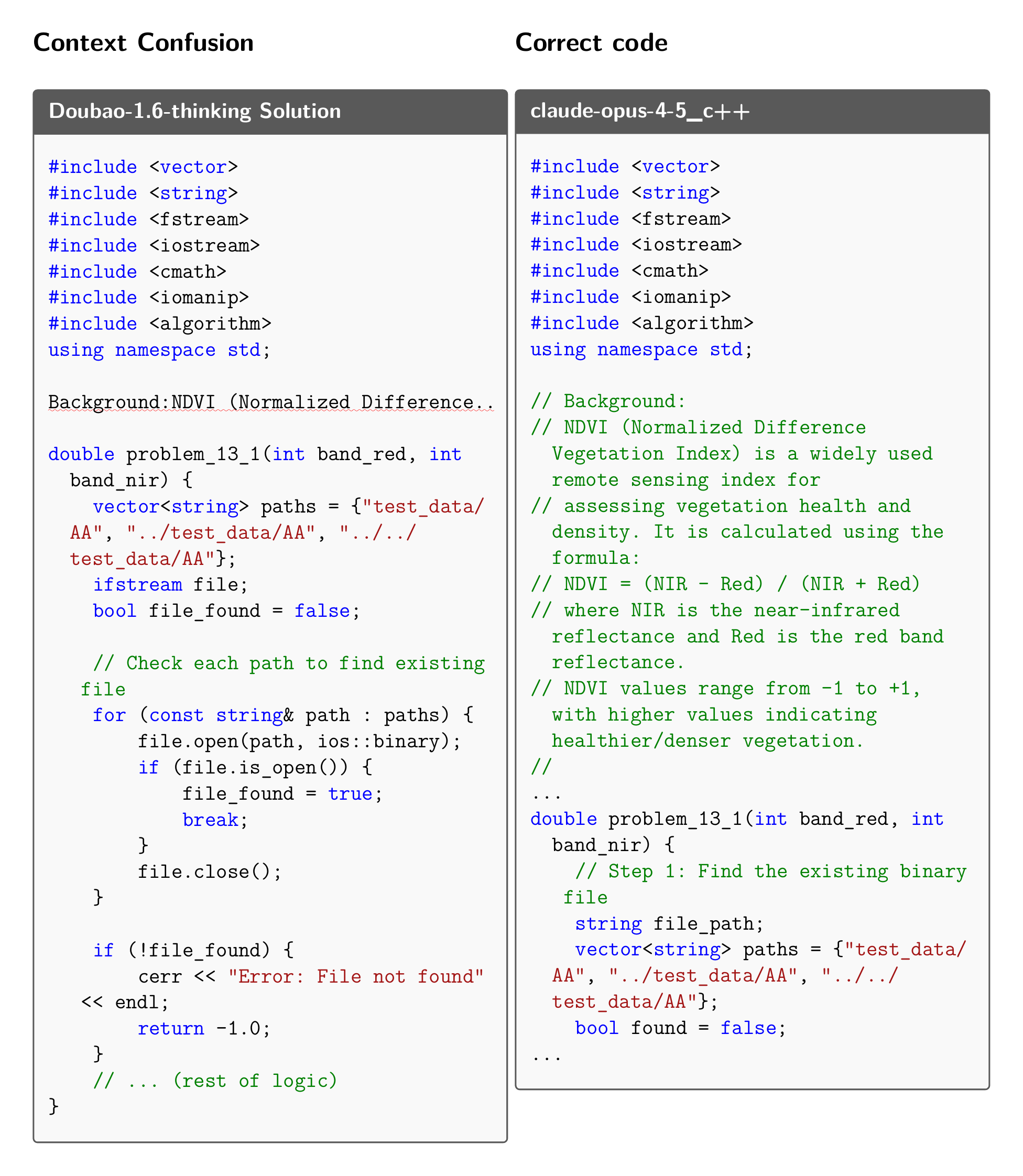}
    \caption{\textbf{Comparative Analysis of Context Segregation.} Left: Doubao (Leaking). Right: Claude (Isolated).}
    \label{fig:compare_2}
\end{figure}

\textbf{Analysis of Claude's Performance:}
This case highlights the superior latent space segregation of Claude. When presented with background theory (NDVI formulas), Doubao fails to inhibit the semantic text from entering the syntactic generation stream, resulting in compilation errors. Claude, conversely, employs a robust information gating mechanism. It effectively distinguishes between contextual knowledge (intended for comments or logic) and executable syntax, ensuring that the background description from the prompt never breaches the code boundary.

\subsubsection{Case 3: Syntax Error: Incomplete Typel Completeness}
\begin{figure}[H]
    \centering
    \SafeIncludeGraphics[width=0.95\textwidth]{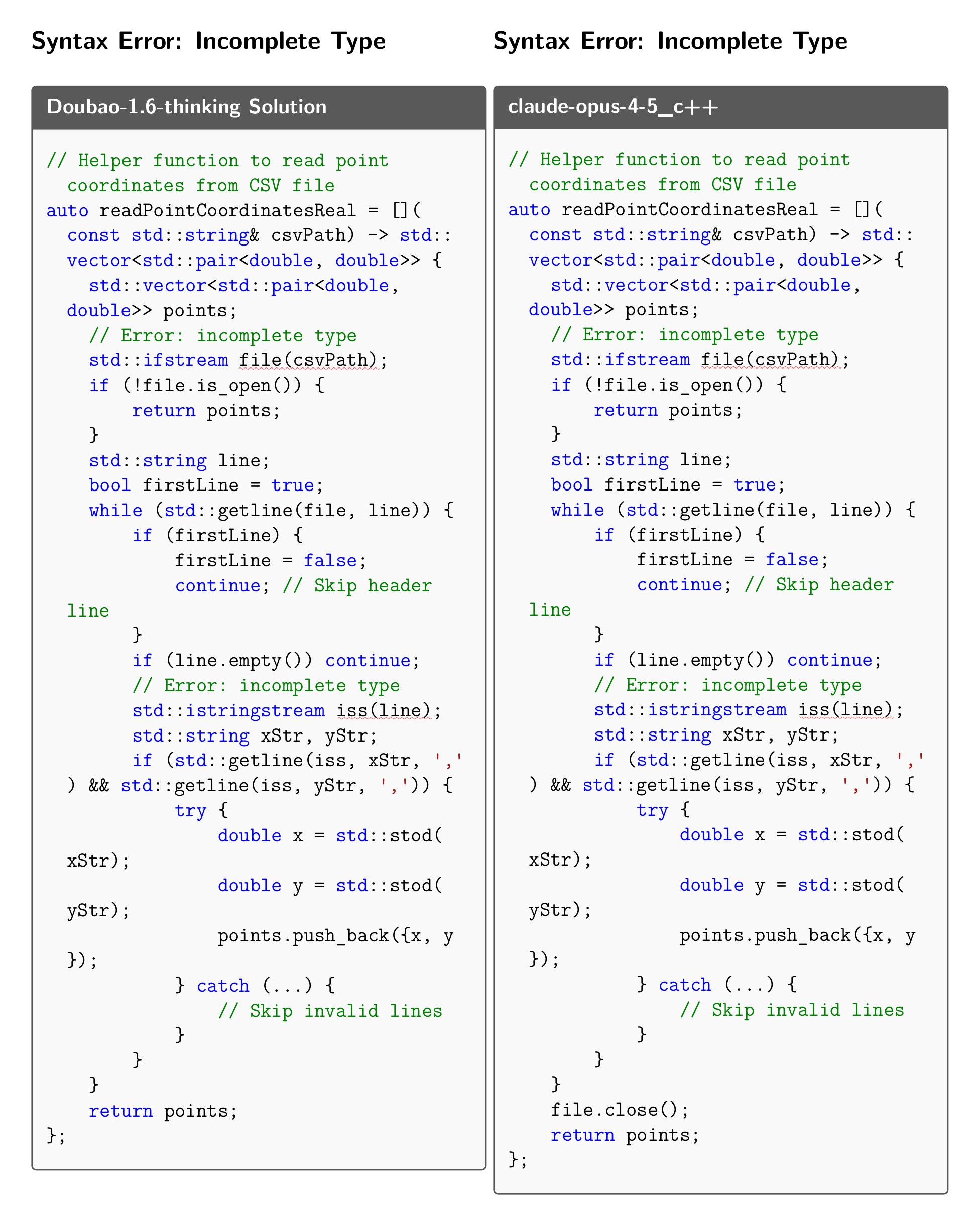}
    \caption{\textbf{Comparative Analysis of Completion Stability.} Left: Doubao (Truncated). Right: Claude (Complete).}
    \label{fig:compare_3}
\end{figure}

\textbf{Analysis of Shared Failure:}
Case 3 reveals a shared vulnerability. Both models exhibit an incomplete type error (specifically regarding \texttt{std::ifstream} usage in this specific context). While Claude is typically robust, this failure suggests that under specific conditions of high cognitive load or ambiguous library dependencies, even leading models can suffer from structural hallucination. Both models appear to assume a type definition exists in the implicit context that was not formally declared, indicating a limitation in symbol table tracking within the attention mechanism for both architectures.

\subsubsection{Case 4: Syntax Error: Redefinition}
\begin{figure}[H]
    \centering
    \SafeIncludeGraphics[width=0.95\textwidth]{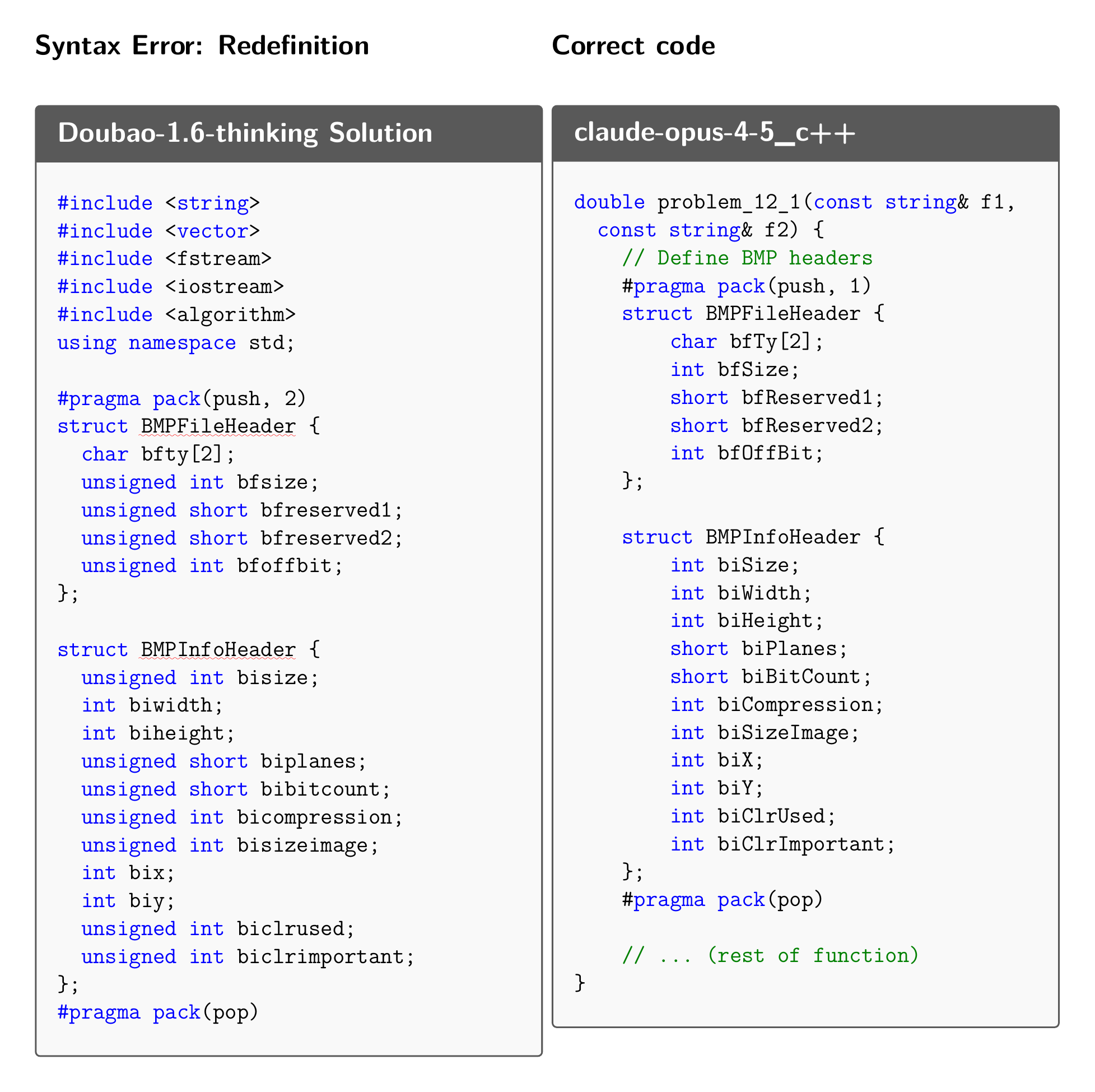}
    \caption{\textbf{Comparative Analysis of Scope Management.} Left: Doubao (Redefinition). Right: Claude (Inner Definition).}
    \label{fig:compare_4}
\end{figure}

\textbf{Mechanistic Analysis:}
Both models struggle with global state tracking, though the manifestations differ:
\begin{itemize}
    \item \textbf{Doubao (Fatal Error):} Commits a compilation error by redefining the \texttt{BMPFileHeader} struct in the global scope. This indicates a failure in the Key-Value (KV) Cache to inhibit the regeneration of tokens that represent already-defined concepts, violating the One Definition Rule.
    
    \item \textbf{Analysis of Claude's Performance:} In this scenario, Doubao commits a fatal compilation error by attempting to redefine the \texttt{BMPInfoHeader} struct globally, violating the One Definition Rule (ODR). Claude avoids this through a defensive architectural heuristic. As seen in the right panel, Claude defines the structure \textit{inside} the function body. While this is arguably an architectural compromise (limiting the reusability of the struct), it ensures compilation success by creating a local scope that cannot conflict with global definitions. Claude prioritizes executability over architectural purity, demonstrating a fail-safe generation strategy.
\end{itemize}

\subsubsection{The Mechanism of Superior Performance: Why Claude Outperforms}
\label{sec:claude_superiority}

The comparative analysis above suggests that the leadership of Claude in code generation is not merely a function of parameter scale, but stems from three fundamental architectural advantages that align high-level reasoning with low-level syntax.

\begin{enumerate}
    \item \textbf{Precise Contextual Segregation (Heuristic Control):} 
    The critical differentiator is the capacity of Claude to maintain strict segregation between generalized reasoning priors and specific domain constraints. In reasoning-enhanced models, the generation of intermediate thoughts activates broad engineering heuristics that can conflict with local requirements. Claude effectively isolates these generalized priors, preventing the phenomenon of Cognitive Over-correction. This ensures that the model adheres to specific legacy specifications (such as signed integers in BMP headers) rather than actively substituting them with incompatible modern conventions found in the training data.
    
    \item \textbf{Long-Horizon State Persistence (KV Cache Fidelity):} 
    Code generation is a state-dependent task; the model must remember a variable defined at token $t=50$ when generating token $t=2000$. The redefinition errors in Doubao (Case 4) indicate a decay in its Key-Value (KV) cache fidelity over time—it fails to retain the information that a structure was already defined. Claude maintains state persistence far more effectively, essentially acting as if it has a larger, more stable working memory for symbol tables, ensuring strict adherence to the One Definition Rule (ODR).
    
    \item \textbf{The Syntax-First Alignment Hypothesis:} 
    We hypothesize that Claude utilizes a dual-stack alignment strategy. While comparison models allow the semantic intent to occasionally displace syntactic requirements, Claude enforces a strict hierarchy where grammatical correctness acts as a hard constraint. Even when the reasoning is complex or the architecture is defensive (as in the local definition in Case 4), the syntax engine ensures the output is valid C++. This prioritization of compilation success over semantic expressiveness is a hallmark of its engineering superiority.
\end{enumerate}

\end{document}